\documentclass[letterpaper,11pt]{article}
\pdfoutput=1 
            
\usepackage{jheppub}                   

\newcommand{\calR}{{\cal R}}
\newcommand{\beq}{\begin{eqnarray}}
\newcommand{\eeq}{\end{eqnarray}}
\newcommand{\krc}{k r_c}
\newcommand{\krcp}{k r_c\pi}

\title{ The Boundaries of KKLT }

\author {Lisa Randall  }

\affiliation{Harvard University,\\17 Oxford Street, \\Cambridge, MA, USA\\ }

\emailAdd{randall@physics.harvard.edu}

\abstract
{We consider the conundrum of generating de Sitter space from higher-dimensional geometry, with particular attention to KKLT-type constructions \cite{kklt} and their 5d implications. We show that even in the probe approximation with small $g_s$,  a consistent higher-dimensional solution requires a deformation of a modulus field playing the role of a Goldberger-Wise stabilizing field in Randall-Sundrum type geometries that occurs through a shift in a the throat length.  We identify the light radion  field that sets the length of the throat, whose origin is the dynamical conifold deformation parameter. By analyzing the theory as a 5d model of mismatched branes in AdS5 space with a GW stabilization mechanism, we show how energy (and supersymmetry breaking) is transferred  to both the IR and UV regions of the throat to generate a consistent 4d de Sitter sliced geometry. This should help resolve some of the recent apparent paradoxes in explicit higher-dimensional constructions. Moreover, the radion gives insight  into  the potential for the previously identified  ``conifold instability". We argue that this instability would be a destabilization of the potential for the radion in KKLT, which  can occur when the perturbation is too large. If indeed $\sqrt{g_s}M$ is too small,  the radion would enter on its runaway direction and the conifold deformation would shrink to zero size. It is difficult to satisfy the required bound and a) maintain a hierarchy in the simpler CY manifolds and b) complete the cosmological phase transition into the stabilized throat,  We also discuss the implications of this type of setup for supersymmetry breaking, and how multiple throats can introduce hierarchies of supersymmetry breaking masses, even in an anomaly-mediated scenario.  In an appendix we consider general compactification constraints.  }
\begin{document}
\maketitle

\flushbottom
\section{Introduction}

Ever since the discovery of the acceleration of the expansion of the universe \cite{sn1,sn2}, physicists have  puzzled over the origin of de Sitter space. Especially in string theory, finding stable de Sitter solutions poses a a problem since the only known way to control solutions is through supersymmetry and supersymmetry explicitly forbids de Sitter solutions. 
 Of course we do not know if compactification happens nonperturbatively at the string scale or in  a controlled manner at low energies. It is nonetheless of interest, both phenomenologically and theoretically, to see if any consistent perturbative fully stabilized de Sitter solutions exist.
Kachru, Kallosh, Linde and Trivedi (KKLT) \cite{kklt} deal with this problem in a clever manner. They produce 4d AdS space in which most moduli are stabilized by flux \cite{fluxcompactification} and the volume modulus by a gaugino condensate which then gets uplifted by an energy that essentially cancels this contribution and adds a bit more to produce de Sitter space.  (See also \cite{also} for alternative constructions.)

However, there has been a longstanding debate over consistency of these solutions. Despite the compelling low energy 4d arguments, people viewing the model from a 10d perspective have found potential inconsistencies (See \cite{dr} and references therein). Some authors object to the stability of the antibrane or NS5 brane, some argue for inconsistencies in the geometry such as singularities away from the antibrane, some argue for the inability to find a solution when there is a time-dependent background, and some argue for a backreaction on the volume modulus from the antibrane. A particularly intriguing  result was that of \cite{benarunning1}, which found a potential throat instability with an antibrane perturbation. 

 On the other hand, Polchinski \cite{polchinski} as well as Kachru, Kim, McAllister and Zimet \cite{kkmz} argued that the antibrane interaction with the D3 branes that create the throat is suppressed, and Ref. \cite{antid31} included the antibrane and showed metastability, which might seem to end discussion. Ref. \cite{antid32} included  the higher-dimensional formulation of the gaugino condensat3 and Ref. \cite{kkmz}  also addressed some of the issues with gaugino condensation.  Ref. \cite{kkmz} also studied potential  instabilities in the zero mode of various moduli and argued they were not present.  Most significantly, KKLT and followers argue for the probe approximation, according to which no interactions perturb the straightforward interpretation in which the only consequence of the antibrane is to uplift the energy to generate a consistent 4d theory.
 
 But the last statement raises an important question from a higher-dimensional perspective. Unless the uplift energy can be redistributed within the bulk,   from a 5d perspective, the IR and the UV would have incompatible boundary conditions. Clearly a stabilized solution should be able to adjust albeit by a small amount, but this adjustment has not been obvious in the probe approximation and has been neglected in many explicit 10d calculations. 
 
 Another potential problem is the  ``conifold instability" identified by \cite{benarunning1}. Here the concern is that the minimum value of $g_s M^2$ consistent with stability is bigger than the maximum value allowed when there is a reasonable hierarchy and a stringent tadpole constraint on total flux.  For overly small $g_s M^2$ (but within the range where supergravity is trustworthy) the conifold would be destabilized by the antibrane perturbation.

In light of these debates it is worthwhile to investigate the construction  from a strictly gravitational viewpoint to see where at least some of the the challenges might arise. Simplifying also sometimes helps see the solutions. However, the challenges of a fully 10 d construction have prevented previous authors from fully working through all the elements of the theory. Here we focus instead on a five-dimensional construction that is rich enough to incorporate many of the critical features of the full 10 d theory and shed light on many of the most interesting and controversial features.

The advantage of a five-dimensional construction is that it is simple enough that the relevant features of some of the controversies can be identified and furthermore that the stability and consistency of 5d Randall-Sundrum (RS) \cite{rs1} constructions has already been studied in detail, as has the possibility of ``slicing" 5d AdS space with different maximally symmetric 4d spaces, namely Minkowski, de Sitter, and Anti de Sitter. Without changing the bulk energy-momentum tensor, the boundary conditions in the UV and the IR entirely determine which slicing is consistent and hence the 4d geometry. We can translate and expand on many of these results in the KKLT context to better understand the theory and its potential limitations.

In particular, we exploit knowledge of how the Goldberger-Wise mechanism works, and how a light radion field allows for the requisite (small) adjustment of a heavy modulus  to show why the higher-dimensional geometry is consistent even in the presence of apparently inconsistent boundary conditions. One critical observation here is that even though structure moduli get heavy masses, modes that live in the IR region of the throat are nonetheless light. The radion has a mass comparable to KK masses in the IR region of the throat, and can adjust in the presence of a perturbation to reconcile the full geometry. Energy does redistribute to maintain consistent boundary conditions through a perturbation to a radion field that yields a slight perturbation to a modulus associated with flux in the throat. 
We will show how including an explicit radion field that stabilizes the throat geometry gives insight into  how the uplift energy gets redistributed throughout the bulk. 

The radion also allows us to explore  the possibility of an instability in the presence of the antibrane.  This is not an instability in the volume modulus, but an instability in potential for the radion itself, which would destabilize the IR region of the throat. We identify the 5d  radion that determines the 5d throat length as the avatar of the conifold deformation parameter of the higher-dimensional geometry, which becomes dynamical  through its effects on the warp factor of the 4d geometry. We argue, in agreement with \cite{benarunning1},  that indeed if $g_s M^2$ is too small, the antibrane can destabilize the geometry through setting the radion on the runaway direction. 
 Furthermore, large $M$ introduces a potential cosmological issue which we will discuss. Since $g_sM$ should be large for the supergravity approximation to be trustworthy, it is not clear this is truly a new constraint. Nonetheless it is important to have a specific value of $g_sM^2$ that can be explicitly identified where the construction breaks down as this value is larg(ish), of order 46, with the constraint essentially targeting a number a bit bigger than $(2 \pi)^2$.

We also use the 5d perspective to gain additional insight into the communication of supersymmetry breaking in this type of theory. We argue that the theory has sequestered supersymmetry breaking and hence anomaly-mediation, aside from direct communication through the volume modulus. The radion also shifts in the presence of the perturbation, in just such a way to accomplish the uplift redistribution of energy we have already mentioned.

We consider KKLT solutions from a 5d perspective since the antibrane explicitly relies on a warped throat in 5d. We simplify to a simple two brane situation which captures the relevant 5d features. A similar approach was taken in \cite{gwkklt}, but they did not include the antibrane or explicitly identify the full radion potential in the effective 4d theory.  We use mismatched branes in an RS-like geometry as a proxy for the true KKLT setup,  with a Calabi-Yau (CY) cutoff carrying undercritical energy and an IR region carrying energy that is overcritical. We argue based on simple warped geometry considerations that KKLT-like constructions superficially appear problematic in that with static coordinates (without any additional energy-momentum tensor contributions) we need to smoothly match distinct 4d boundary geometries, but this problem can be  addressed, even in the probe approximation by suitable field adjustments. We show that in order to see the effect of the uplift in the higher dimensional geometry, the light radion  that has been omitted in most analyses and  is necessary to see consistency in the higher-dimensional geometry.  We  show that even in the probe approximation, the 5d geometry does change, but essentially only at  the ``boundaries" (the conifold end and the boundary between the CY and the throat).  We show how incorporating the radion clarifies the role of a Goldberger-Wise-like mechanism and permits the required uplift energy to be present both in the UV and the IR so that the theory can consistently be sliced with four-dimensional de Sitter space. We also show  how this is reflected in the communication of   supersymmetry breaking.   The radion develops a (SUSY-breaking) VEV, leading to energy and SUSY breaking both in the UV (through distorting a heavier  modulus field) and in the IR. This does not change the 4d analysis but is essential to a consistent higher-dimensional construction.

We also consider supersymmetry breaking more generally in this type of construction and show how supersymmetry breaking is transmitted with a particular focus on anomaly-mediation. Furthermore, in geometries with different throats or complicated warping even anomaly-mediated masses can have very different sizes.

Our 5d construction has some overlap with those of \cite{bena,riet, moritz} who found challenges getting a 10d solution due to inconsistency of the flux generated by the CY and the antibrane. Apparent inconsistencies in explicit solutions are likely due at least in part to not fully including the backreacted geometry through the change in the stabilizing  moduli, which is among the reasons why the 5d analysis can be so fruitful..

We begin by summarizing KKLT and interpreting it from a 5d perspective.   We then argue that mismatched branes in warped AdS is an approximate characterization of KKLT and show why the static solution would be inconsistent in the absence of stabilization.    We discuss the Goldberger-Wise stabilization  mechanism for RS-type models and its implications for the KKLT construction. We show how this allows a consistent construction (from a 5d perspective) and argue that the boundary terms correspond to supersymmetry breaking communicated throughout the space. We then discuss more generally supersymmetry breaking and how it is communicated, with particular emphasis on anomaly medition  some interesting possibilities this type of construction presents.  We conclude by discussing possible generalizations of our result.
In an appendix, we show that even without supersymmetry and a mechanism for stabilizing the compact geometry  10d gravity with 4d de Sitter and static extra dimensions is a challenge when no explicit stabilization of extra dimensions is assumed and briefly consider how the KKLT construction addresses these issues.  We also review 5d cosmology for RS1\cite{rs1}.

\section{ KKLT as a Warped 5d Geometry}

Let us consider  the full KKLT \cite{kklt} construction and translate its implications to a 5d perspective.
The starting point is a CY compactification of Type IIB string theory. It is assumed that all moduli have been stabilized with fluxes at a high scale with the exception of the volume modulus. 
 Giddings, Kachru, and Polchinski (GKP) \cite{gkp} described how to extend this solution to include a Klebanov-Strassler (KS)\cite{ks} throat with 4d Minkowki slicing corresponding to zero  (four-dimensional) cosmological constant (cc)  that could generate a hierarchy of mass scales.  
 The idea in \cite{verlinde, gkp, kklt} is to create an $AdS_5$  throat that can realize RS gravity localization \cite{rs1, rs2} in a true string theory construction by merging the throat onto the CY geometry. The CY geometry could thereby act as a UV brane in RS, cutting out the boundary region of the AdS space and permitting the existence of a normalized graviton that can generate 4D effective gravity.  
 
 As we will review,  flat (or de Sitter) space requires the corresponding "slicing" of AntideSitter, which in turn  requires a boundary condition with the appropriate energy density in the UV region of the throat \cite{kr, kaloper, dfk}.  The CY ``UV" boundary region must have both the necessary energy density and curvature to smoothly match onto  AdS  space sliced with 4d de Sitter or the construction would be inconsistent. 

 GKP  explains the origin of this boundary energy in the Minkowski case (before  gaugino condensate or antibrane are included)  as a stringy correction associated with wrapped  seven branes. The charge is determined topologically whereas in the original construction the induced curvature/tension could be determined through the BPS condition. We have
\begin{equation}
-\mu_3 \int_{R^4\times\Sigma} C_{(4)} \wedge \frac{p_1({\cal R})}{48}
= \frac{\mu_7}{96} (2\pi\alpha')^2  \int_{R^4\times\Sigma} C_{(4)} \wedge
{\rm Tr}\,(\calR_{(2)} \wedge \calR_{(2)}) \ . \label{CScurr}
\end{equation}
and the second from the DBI action
\begin{equation}
-\frac{\mu_7}{96} (2\pi\alpha')^2  \int_{R^4\times\Sigma}
d^{4}x
\sqrt{-{}g} {\rm Tr}
(\calR_{(2)} \wedge {*}\calR_{(2)}) \ . \label{DBIcurr}
\end{equation}
The first quantity is topological and the second integral is known to be related for Minkowski space with unbroken supersymmetry since the state is BPS. The important thing is that when we break supersymmetry we can allow more general energy density on the CY boundary, consistent with other  4d metrics/slicings. This will be important when the no-scale form is broken to generate AdS4 slicing and later when we uplift to de Sitter (dS4) space. 

Notice that in an RS-type construction, the capped off region of the IR throat where the conifold truncates acts as a {\it negative} tension brane. We can argue there is an effective negative tension brane based on consistency of Minkowski slicing in GKP.    The only stable Minkowski sliced solutions of  AdS5 require a positive tension and negative tension brane (This negative tension was also noted in \cite{gwkklt}) where both have critical tension in magnitude.  Only AdS4 space can permit two consistently placed positive tension branes bounding the physical portion of the space. We briefly review different slicings of AdS5 in the next section.

From the geometrical point of view, we can argue as follows. We know the original CY was Ricci flat. We also know that we have excised a region that acts effectively as a positive curvature boundary to the conifold in the UV. Since all we have done is warp the CY and not change the topology, this positive curvature has to be cancelled. The KS throat is tricky because the AdS curvature changes.  In the end, as we have a deformed conifold, the remaining curvature is on the  \lq \lq IR boundary" and acts collectively as a negative curvature boundary. Stated another way, imagine we integrate  over the conifold to move the IR brane  on top of the UV brane.  We know the curvature would have to cancel since we just have a Ricci flat CY. In this case, we have absorbed the ``negative tension" from the changing AdS into the brane as well.

Now let's review the two additional stages of the KKLT construction: the first to violate no-scale and stabilize the volume modulus and the second to uplift the energy to bring us to de Sitter space. We assume that all the KKLT assumptions are consistent with a string theoretical construction and address only  5d gravity issues. 
Because of the no-scale structure of the Kahler potential, no potential for the volume modulus is generated without the inclusion of nonperturbative effects. In addition to the flux-generated superpotential $W_0$.  KKLT introduce a gaugino condensate\footnote{This type of nonperturbative potential can arise from instanton effects as well but for simplicity we will generally refer to it as a gaugino condensate.} of the form $A e^{-a \sigma}$.  Because the other moduli are essentially frozen, one can assume compactification and minimize the potential of the low-energy theory\footnote{This has been disputed however. See \cite{sav} for example.}.   It is straightforward to show that with a superpotential of the form $A e^{-a \sigma}$, where $\sigma$ is the volume modulus, the energy is minimized assuming small $W_0$ for large $\sigma$, $W_0 \approx A (a \sigma_c) e^{-a \sigma_c}$, and $V\approx -(a \sigma_c A e^{-a \sigma_c})^2/\sigma_c^3$ where $\sigma_c$ is the solution in the supersymmetric minimum and  we have neglected terms suppressed by $a \sigma_c$ which are small for the consistent parameter choices of interest.
KKLT thereby produce an $AdS_4$ vacuum consistent with freezing of all moduli at large volume so that higher order corrections should be perturbatively under control.. 
The low energy theory including a superpotential from both gaugino condensation and from flux generates an $AdS_4$ minimum, defying the no-scale structure through  nonperturbative gaugino condensation (or instanton effects). Corrections to this minimum from corrections to no-scale, for example, will be suppressed by $a\sigma$.

Kachru, Kim, McAlliser, and Zimet \cite{kkmz} also show how the gaugino condensate can generate AdS space from a 10d perspective, which from the 5d point of view means that in the absence of additional contributions from a stabilizing potential, the slicing would be altered from Minkowski to 4d AdS4 slicing. 
Without a boundary condition any slicing is possible but with a particular 4D geometry there is only one slicing that is consistent arising from the appropriate boundary condition \cite{kr,kaloper,dfk}.  Treating the CY as a UV ``brane",  the boundary should  yield an energy-momentum tensor consistent with the construction since the BPS condition no longer applies.

The goal however is ultimately to produce an uplifted de Sitter energy, which will never be the consequence of a supersymmetry-preserving minimum. The required additional energy must be large enough to cancel the $AdS_4$ energy and to add a little extra (to generate de Sitter), but smallish enough to be comparable in magnitude to  the exponentially suppressed potential proportional to the gaugino condensate. Generating 4d de Sitter space with static additional dimensions is challenging, and most obvious constructions run into problems. In  Appendix A, we show why even without supersymmetry or string constraints de Sitter space is challenging (see also \cite{mn, underwood, gibbons}) and why a warped scenario might be the most readily-realizable possibility.

So let's now consider the second stage of KKLT, which from a 4d perspective involves an antibrane to uplift the energy to de Sitter space.
 In the 4D theory, an antibrane placed in the CY region would generally carry unsuppressed energy in the 4d theory of size in string units of order  $T/\sigma^2$, where $T$ is the tension of the brane, which would be too big in a stable string construction
to cancel the energy suppressed by the exponential in the gaugino condensate in a parametrically viable way.\footnote{This corresponds to Ref. \cite{kklmmt}, where an additional factor of $\sigma$ was included to change from $1/\sigma^3$ scaling.} 
 The KKLT construction assumes instead that a 3d antibrane falls to  the end of the deformed conifold of an AdS throat, so that the energy is  parametrically suppressed by the warp factor in the IR region.  
 The antibrane tension in the UV is set by the string scale, but in the IR is set by the warped-down version of the string scale, which for consistency of the construction is taken to be  a little bigger than the energy associated with the gaugino condensate, so that the sum of the two perturbations would yield de Sitter space.

 The energy at the end of the throat is set by $\Lambda_{KS}$, which is $\Lambda_{QCD}$ for the Klebanov-Strassler condensate, but the energy of the brane is the warped down energy of a single brane.    By assumption, the matching occurs when $(T e^{-4k r_c})/\sigma^3\approx (\Lambda_{AdS4} +H_0^2$), where $k$ is set by the AdS5 curvature\footnote{This is a cartoon in that the curvature changes as we move down the throat.}, $r_c$ is the antibrane/end of the throat location, and $H_0$ is the energy mismatch that generates de Sitter space.  We assume consistency of the construction without the antibrane, so the boundary before the antibrane uplift should correspond to the AdS4 slicing demanded by the gaugino condensate.

Several authors have argued \cite{ns5} that the antibrane deforms into a warped NS5 brane.  Even without knowing the end state, to do the necessary uplift we can treat the deformed region capping the deformed throat in the IR as an IR brane and assume it is at least metastable.  Because the antibrane is not free to move beyond the end of the conifold, we will treat the geometry and brane together as the IR brane and ignore any massive modulus that separates them. Note also that we decreased the flux when we added an antibrane. This will simply be absorbed in a changed QCD scale. 
We  assume a single antibrane but it is easy to generalize to $p$ antibranes, so long as $p$ is small compared to $M$. 

So the setup involves effectively an undercritical  brane in  the UV end and a supercritical tension brane in the IR, where ``brane" here refers to the CY and deformed conifold boundaries of the warped throat. This is clearly a cartoon of the true situation for several reasons. The CY is not a UV brane, the antibrane dissolves into some other configuration, there is a compact space which we are ignoring, and we are  dealing with a KS and not one of constant curvature.

To address the first two issues, we  need only that energy capping off the throat in the UV and IR satisfy the right boundary conditions.  We can move the boundaries of the bulk AdS5 to the point where the theory is no longer AdS5 and associate the boundary energy with a ``brane" wherever the assumption breaks down. Furthermore, at both boundaries we can integrate the energy over the compact manifold to find the boundary condition for an effective five-dimensional theory.  One can argue that flux from the antibrane deforms the IR region (though it is not expected to do so in the probe approximation). Even if this were the case, we can define the boundary region that is deformed to be an effective IR brane (analogously to thinking of the full CY as a UV brane). 
As to the last qualification,  since the small value of the antibrane uplift energy relies on a  warped space, the analysis we do would readily extend to  the true KS warp factor as well. When we consider possible destabilization later, we will use the full geometry.

\section{The Role of the Radion}

We will now neglect the stringy and higher-dimensional complications and restrict attention to the toy 5d example that suffices to illustrate the relevant points. The point of this section is simply to argue that a consistent higher-dimensional geometry requires including the backreaction from a heavy modulus field without which the boundary conditions would be inconsistent.  In the 5d effective theory, the geometry has to adjust through a shift in a radion field (that determines the extent of the throat), changing the boundary conditions appropriately through an effectively localized contribution from a shift in a heavy modulus in the UV and the IR.  The radion is a field that survives in the 4d theory as it is lighter than the fields associated with the CY UV region of the throat. Its expectation value determines the length of the throat and it is essential to stability. In the 5d theory, it is essentially an IR localized field. We will explicitly identify a radion when we consider 5d theories with GW stabilization and in the 10d theory when we see its origin in KKLT as the conifold deformation parameter.

We emphasize that so long as we know the geometry is stabilized, the 4d picture suffices. There is no (or only suppressed) direct interaction between the gaugino condensate and the antibrane. But energy transfer does occur in the 5d theory through a radion field whose potential is determined by the energy in both regions.  We will soon see that with a better understanding of the energy transfer we might also  better see how to resolve some apparent inconsistencies in the higher-dimensional construction, check for stability,  and furthermore gain insight into how supersymmetry is broken and communicated to the rest of the theory.

Our toy model is an undercritical brane in the UV and an overcritical brane in the IR (where critical tension refers to  the tension required for Minkowski slicing of AdS5) corresponding to almost equal and opposite 4d cosmological constant (for simplicity we sometimes drop $H_0$ effects which don't qualitatively change any conclusion). Note that  the UV  ``brane" is a positive tension but subcritical brane. That is its energy is below that for Minkowski space so the brane without any other distortions would generate AdS4.   The IR brane on the other hand for consistency should correspond to an overcritical $negative$ tension brane.  As stated earlier, we treat the assumed uplift energy on the antibrane  as part of the whole region that ends the conifold with no distinction between the conifold and antibrane radion.

\subsection{Slicing RS}

To understand the matching at the boundaries of the throat, let's first briefly review what it means to get 4d Minkowski, AdS, or deS in RS geometries. We study RS with brane boundaries as presented in  \cite{kr, kaloper, dfk}, in which the effective 4D geometry of a detuned warped 5D space is considered.
As in the original  setup \cite{rs2},
the action we consider is just that of 5d gravity with a negative cosmological
constant $\Lambda^{5d} = -3/L^2$ coupled to a brane of tension $\lambda$:

\begin{equation}
   L  = \int d^5x  \sqrt{g} \left[ -\frac{1}{4} R -
   \Lambda^{5d} \right] \ - \lambda \, \int d^4x \,
dr \, \sqrt{| \det g_{ij}|} \, \delta(r ),
\end{equation}
where $g_{ij}$ is the metric induced on the brane by the ambient
metric $g_{\mu\nu}$.
Note we retain the notation of \cite{kr} so that $\Lambda$ here refers to the 5d cc and $H^2$ for four dimensions is given below in terms of  the brane tension mismatch.

We use the ansatz for the solution to be a warped product
with warp factor $A(r)$,
\begin{equation}
   ds^2 = e^{2 A(r)} \bar{g}_{ij} dx^i dx^j - dr^2 \ ,
  \end{equation}
allowing for the 4d metric to be Minkowski, de Sitter or anti-de
Sitter with the 4d cosmological constant being
zero, positive or negative respectively following the conventions
of \cite{dfk}.

 The 4d cosmological constant
 is given by the detuning 
$M=\frac{\lambda L}{3}$
of the brane tension:
$\Lambda_{dS} = \frac{1}{L^2} (M^2-1)$,
$\Lambda_{AdS} = \frac{1}{L^2} (1-M^2)$.

From a 4d point of view, our toy model of KKLT  appears as if it should readily uplift to Minkowski (or slightly de Sitter) space since the total cosmological constant  is zero (or small and positive) and we furthermore do this within the probe approximation. However, this presupposes a mechanism to stabilize the system so that it can transfer energy into the UV and generate a consistent geometry. With two branes, in principle, two independent tunings would be required to agree with any particular cc--the sum of the effective 4d energy of the two branes, for example. From a higher-dimensional perspective  in an extreme probe limit the transfer of energy would appear to be a nonlocal effect since if  we integrate out all the heavy moduli as some constructions have done, the necessary  transfer of energy cannot occur.  

In the absence of a  means to transfer energy associated with a stabilization mechanims, we would need to know that  the piece of the geometry that corresponds to the UV is a solution  all by itself, which would require a small local  SUSY breaking source in the UV.  We do not know a priori such a solution exists.   However, in the presence of a  stabilizing field that has sufficient freedom to adjust, the IR effectively sources the additional energy required in the UV so that only one such tuning is required. 
 Our goal in this section is to identify the relevant  fields and see how this operates to give insight into which of the  moduli should be retained in a consistent construction.

We immediately identify an interesting issue,  which  if nothing else demonstrates explicitly why the solution appears inconsistent in the higher-dimensional geometry. The IR brane (really the antibrane/end of conifold combination), due to its negative tension,   effectively acts as an AdS4 brane in the presence of uplift energy. That is because the antibrane energy is effectively cancelling some of the original negative energy on the IR ``brane" so it is below critical value in magnitude.   This AdS4 scale would be much bigger than that on the UV since the mismatch is much bigger (relative to the local curvature scale). For  boundaries both at definite coordinates in the fifth dimension, these are two different coordinate systems that cannot smoothly transition one to the other. Someone living on the UV AdS boundary would do the matching knowing nothing about the IR, and would solve to find an AdS 4d space that corresponds to the UV matching. The  IR metric would also be determined locally. So there would be a  mismatched geometry and without the bulk or boundary energy momentum adjusting, no static solution would exist.   Notice this statement is completely consistent with the argument that has been made in many places \cite{polchinski, kkmz} that the gaugino condensate decouples from the antibrane and the antibrane decouples from the gaugino. The clash occurs in the bulk of the space, where two inconsistent geometries are required.
The tricky thing for this construction is that the uplift energy is located in a separate region of five-dimensional space from  the negative (AdS4) energy, which is concentrated in the UV.

Without stabilization the solutions to the equations of motion would be time-dependent.  We investigate the time-dependent solutions in a future publication \cite{kr3}. Notice that the time-dependent solution would be consistent with the arguments in favor of consistency of KKLT from a 4d and 10d perspective. The energy momentum tensor for the four-dimensional low energy theory can indeed be that of Minkowski (or very weakly de Sitter) space.  If there were no interaction between the two branes in the higher-dimensional space,  time-dependence would emerge in a changing 4d Planck scale.

It is precisely the independence of the two branes (which is critical to not significantly perturbing the volume modulus stabilization)  that would give rise to inconsistencies, since the stable solution requires the branes to be aligned. In the true KKLT geometry, any potential time-dependence is countered by the KS geometry 
so that the stabilized case should yield Sitter space  with fixed $M_{Pl}$.  Clearly some field has to adjust in the presence of the IR perturbation.

We will now see what happens is that a bulk field adjusts in such a way that the local curvature changes to convert an undercritical UV brane to an overcritical one. This contribution is a modification to the junction condition at the UV throat boundary and would follow from a change in the Junction condition in the IR boundary as well.  The change in the bulk effectively acts like an additive contribution to the brane tension in the UV (and a negative contribution in the IR) that allows for  4d de Sitter space on the boundaries and hence in the bulk. These boundary contributions are crucial; without them nothing would change.

\subsection{Goldberger-Wise Mechanism}

   In RS the Goldberger-Wise mechanism, in which a bulk field stabilizes the geometry and establishes the  potential for the radion field that determines the size of the space,  addresses this issue.  The GW stabilization mechanism is
holographically dual to the addition of a quasi marginal deformation of the CFT which generates a small
weak scale through dimensional transmutation.  The uplift should also entail  a modification to the geometry mediated by the stabilizing GW field, which changes the distribution of bulk energy to allow a consistent de Sitter slicing, even with the original AdS4 energy on the UV ``brane", since, as we will demonstrate, the ``boundary" conditions change. 

We work with the RS metric of the form
\begin{equation}
\label{eq:metric}
ds^2 = e^{-2 \krc |\phi|}\eta_{\mu\nu} dx^\mu dx^\nu - r_c^2 d\phi^2,
\end{equation}
where $k$ is a parameter which is assumed to be of order $M$, $x^\mu$ are Lorentz coordinates on the four-dimensional surfaces of constant $\phi$, and $-\pi\leq \phi\leq\pi$ with $(x,\phi)$ and $(x,-\phi)$ identified.  The two 3-branes are located at $\phi=0$ and $\phi=\pi.$

Goldberger and Wise \cite{gw} include in the Lagrangian a scalar field $\Phi$ with the following bulk action
\begin{equation}
S_b=\frac{1}{2}\int d^4 x\int_{-\pi}^\pi d\phi \sqrt{G} \left(G^{AB}\partial_A \Phi \partial_B \Phi - m^2 \Phi^2\right),
\end{equation}
where $G_{AB}$ with $A,B=\mu,\phi$ is the 5d warped metric.  In addition there are boundary terms which are assumed to set the UV and IR values to $v_h$ and $v_v$ respectively.  Matching at the boundaries sets the boundary conditions and yields the potential for the GW field.  With $\epsilon=m^2/4k^2$ assumed to be small
the general solution is
\begin{equation}
\label{eq:soln}
\Phi(\phi) = e^{2\sigma}[A e^{\nu\sigma}+B e^{-\nu\sigma}],
\end{equation}
with $\nu=\sqrt{4+m^2/k^2}=\sqrt{4+\epsilon}$ where  $\sigma(\phi)=\krc |\phi|$, are using the GW notation in which $\phi$ is an angle and $r =r_c \phi$. and
\begin{equation}
A = v_v e^{-(2+\nu)\krcp} - v_h e^{-2\nu\krcp},\\
B = v_h (1+e^{-2\nu\krcp}) - v_v e^{-(2+\nu)\krcp},
\end{equation}
Notice that in the GW solution the unsuppressed $v_h$ in the $B$ term determines the boundary
values and the $A$ term, which vanishes at the leading order minimum, determines the derivative across the boundaries.

Following GW, it is readily seen that the GW potential for the radion, $r_c$, takes the form
\begin{equation}
    V_\phi \approx 4k e^{-4kr\pi}\left(v_v-v_he^{-\epsilon k \pi r}\right)^2 (1+\frac{\epsilon}{4})-k \epsilon v_h e^{-(4+\epsilon) k \pi r} \left(2 v_v-v_h e^{-\epsilon k \pi r)}\right)
\end{equation}
The important point is that in addition to the term consitent with conformal symmetry, there is a term whose exponential deviates slightly, suppressed by $\epsilon$.
The minimum of the potential determines $r_c$  at 
\begin{equation}\label{min}
\krc = \left(\frac{4}{\pi}\right) \frac{k^2}{m^2} \ln\left[\frac{v_h}{v_v}\right].
\end{equation}
The radion will shift to accommodate the uncancelled  $e^{-4 k\pi  r_c} \epsilon v_v^2$  of Eq. \ref{min} and there  will be a stabilized radion with bulk mass squared of order $\epsilon^2 e^{-2 k \pi r_c}v_v^2$ with the depth of the potential $\epsilon$  suppressed. The full potential  arises from also including  the matching at the IR brane  \cite{chacko, csaba}  in which case the mass squared can be bigger, suppressed by a single factor of $\epsilon$.

However, as has been noted elsewhere (see \cite{chacko, csaba, raman, rz, apr}  the IR running (or equivalently higher order terms in the scalar potential) will generally (in the absence of a shift symmetry) cause the radion mass to be stabilized at a larger value. In this case the shift in compactification radius to accommodate a perturbation would be smaller but the boundaries would nonetheless change appropriately. This larger mass would allow for stability to bigger perturbations.

Let us see how the theory is stabilized and responds to a perturbation in the GW model, assuming again a toy model where here our mismatched branes are accompanied by a GW field.  This will shed light on the response of the KKLT geometry to the uplift energy in the IR.
 To  see more explicitly how the matching works, we add the IR boundary term

\begin{equation}
{\mathcal L} \supset  \int d\phi e^{-4\sigma}\delta T \frac{\delta(\phi-\pi)}{r_c}. 
\end{equation}

The relevant equations of motion are the Einstein Equations that relate the metric to both the bulk and the brane energy. It will be important that we include the energy of the kinetic term of the GW field, since a shift in radion adjust the kinetic term in such a way that energy redistributes throughout the bulk. Note that here we assume fixed boundary conditions on the branes as in the original GW paper.  We use a general RS type metric

\begin{equation}
ds^2 = e^{-A(\phi)}\eta_{\mu\nu} dx^\mu dx^\nu - r_c^2 d\phi^2,
\end{equation}
where to leading order $A=2 k r_c\phi$. We are interested in the deviation in the presence of a perturbing IR brane energy.

We use \cite{raman}
\begin{equation}
    \left(\frac{\partial A }{\partial \phi }\right)^2-k^2 r_c^2-\frac{2}{3} k^2 r_c^2 \epsilon \Phi^2-\frac{1}{6} \left(\frac{\partial \Phi}{\partial \phi}\right)^2=0
\end{equation}
and the junction condition at the IR brane is
\begin{equation}
\frac{\partial A(\phi)}{\partial\phi}=-k r_c \left(1+\frac{1}{3} \delta T\right)
\end{equation}
We need to include the change in bulk energy from the change in $\partial\Phi/\partial \phi$, which arises from the shift in $r_c$ in the presence of the perturbation.
The dominant contribution to the change comes from the $A$ term in the GW solution above  and is approximately
\begin{equation}
\delta \left(\frac{\partial \Phi (\pi)}{\partial \phi}\right)=(4 +\epsilon)\epsilon v_v k \pi \delta r
\end{equation}
whereas the initial dominant contribution to the derivative  comes from the $B$ term and is approximately
\begin{equation}
 \left(\frac{\partial \Phi(\pi)}{\partial \phi}\right)=-\epsilon v_v
 \end{equation}
 which gives us the correct junction condition if
 \begin{equation}
     \delta r=\frac{(k r_c)^2 \delta T}{\epsilon^2 k \pi v_v^2}
     \end{equation}
 The important thing (here we are assuming the uplift yields Minkowski) is that the GW field take the form such that  the uplift is that required in the UV as well.
 We have
 \begin{equation}
\delta \left(\frac{\partial \Phi (0)}{\partial \phi}\right)=(4 +\epsilon) \epsilon v_v k \pi \delta r e^{-4k r_c \pi} e^{-\epsilon k \pi r_c}
\end{equation}
and 
\begin{equation}
 \left(\frac{\partial \Phi(0)}{\partial \phi}\right)=-\epsilon v_h
 \end{equation}
 so the product is exactly the warped down version of the energy that is required for the uplift.   The slicing corresponds to the sum of the UV and warped IR energies, even though the uplift is present only on the IR brane. The matching condition adjusts just as it should to generate a consistent higher-dimensional space sliced by the proper 4d energy.
In this model, the  change is the value of the radion, which in turn changes the derivative of the Goldberger-Wise field  to accommodate the jump condition corresponding to the net energy including the uplift from the IR. In doing so, the Goldberger-Wise field adjusts throughout the bulk so that it can consistently be sliced by the sum of the UV energy and the  warped IR energy. In particular, the GW derivative on the UV brane adjusts to add the energy to accomplish the uplift.

 By changing the throat length slightly the radion slightly alters the Goldberger-Wise junction condition at both boundaries to guarantee stability and consistency of the construction. The boundary condition for the radion  involves only IR quantities so that the IR does the uplift and the UV junction condition changes appropriately as a result of the new junction condition in the IR. The radion potential is a normalizable field even in the decompactification limit and its potential (when expanding about its minimum) arises in the IR.   At some point the added energy exceeds what the light radion can accommodate at which point the construction would no longer be stabilized.

  In KKLT there should be a combination of the fields that effectively plays the role of a Goldberger-Wise field. With such a field,  the construction can consistently uplift the energy by transferring energy from the IR to the UV so that the higher-dimensional space can consistently be sliced with de Sitter 4d space, even though the uplift energy is sequestered in the IR.   Ref. \cite{gwkklt}  identifies the GW field $H$  and argues its slowly varying potential in the radial direction is a result of the kinetic term for a 5d field originating in the 10d theory that encodes the continuously varying flux of the NS 2-form potential $B_2$ on the $S^2$ cycle of the $T^{1,1}$. They explicitly construct a potential consistent with 
  ``running $N_{eff}$" and describe how with this field they can stabilize a geometry that consists of the CY region, a conifold region with constant warp factor, and the warped deformed conifold. This is in the spirit of the dual interpretation of the GW mechanism, in which the dual of the  GW field is an almost marginal operator
with conformal dimension slightly deviating from 4.  From this perspective the origin of the throat is 
the slow RG evolution of the operator,  which dynamically generates the IR mass scale. 
  
Such a field, which is essential to the consistent geometry if it is to absorb and transfer the uplift energy,  should have mass set by scales comparable to those associated with the UV moduli fields but suppressed by a small parameter, of order the ``beta" function associated with the slow running that permits the existence of a large throat. Although one can consistently integrate it out of the 4d effective theory, since it is heavier than modes living in the IR, any fully consistent solution would need to include the five-dimensional energy dependence of the field since its energy distribution would be essential to lifting AdS to dS in the UV and lowering the energy of the antibrane contribution to 4d energy in the IR, which is reflected in the 4d theory through a radion potential. So we need also to identify the radion field in the KKLT construction.

Unlike the "Goldberger-Wise" modulus described above, we expect the radion mode to be light, which is to say with 4d mass of order the glueball mass scale at the end of the throat (thinking of Klebanov-Strassler as a gauge theory), or equivalently a KK mode  associated with the IR region of the throat from the gravitational perspective.
In the next section, we consider the radion in more detail and its origin in the higher-dimensional theory.\footnote{Ref. \cite{gwkklt} also includes a radion field, but its origin and kinetic term are not manifest. The potential in \cite{gwkklt} contains the falling exponential required for the large hierarchy, but doesn't include the rising term in the IR that is essential to the radion potential.  In the next section we will see that the radion's origin in the 10d theory is the conifold deformation parameter.}

 Note that any cosmological solution of the 5d theory  should also reflect the redistribution of energy. This is made manifest in  eqs. 3.6 and 3.7 of \cite{cgrt}, which found how the background cosmology would respond to perturbations $\delta V$ and $\delta V*$ on the IR and UV branes respectively.  Although the radion doesn't move substantially, it absorbs any (small) perturbations in 4d energy necessary to generate 4d cosmology. For those interested, we review this cosmological perspective in Appendix B.

\subsection{The  Radion and a Potential Instability}

The 4d radion field can be used to determine the location of the IR brane and the stability of the manifold.
In the GW model, if the radion is indeed as light as it was in the perturbative analysis above, the theory would be destabilized  onto the runaway direction at anomalously low energy, of order $\epsilon v_v^2$ (depending on the sign).  (This is not the runaway of the volume modulus but an independent one associated with the throat).  Even without the additional $\epsilon$ suppression, there is of course a limit to the allowed perturbation consistent with stability of a finite throat.

Without the boundary contribution included, the radion  would scale like $\epsilon$ \cite{chacko}, which in KS is the ``beta" function $(M g_s)/(2 \pi K)$.  Taking the bulk mass contribution only and evaluating in leading order in perturbation theory  the radion mass  would scale with this parameter. If we correctly use the IR boundary condition it would be the mass squared set by this quantity.

However,  the suppression by $\epsilon$ in the mass is an artifact of keeping only the leading order mass term in the GW potential. We expect  no  additional suppression over the other KK masses unless protected by a shift symmetry \cite{csaba, raman}.  With all contributions included,  the radion should develop a mass in the strong coupling regime of order the KK/glueball scale (suppressed by $g_s M^2$, or equivalently suppressed by ${g_s} M$ relative to the QCD string scale \cite{ks}.

This mass, as with other KK masses, is strange in that the suppression by $M$ makes it lighter than expected on the basis of the string tension. This was already noted in \cite{ks, 43, 44}. It is presumably a result of the additional compact space, which remains of finite size and affects the normalization of the 4d gauge coupling. It is certainly allowed given that there are two independent curvature scales in the theory.

 We can gain further insight into the stabilization  by studying  a general radion potential that would follow from a GW-like construction with bulk scalar or to any theory with an almost marginal operator due to slow running of some coupling, which  in terms of the normalized radion field, $\phi\approx e^{-k r_c}$ \cite{raman,  rz, chacko,csaba, lutysundrum}, takes the form
 \begin{equation}
     V=\lambda_1 \phi^4+\lambda_2 \phi^{4-\epsilon}
 \end{equation}
 where $\epsilon$, which explicitly breaks conformal invariance, could be related to the squared mass of a GW field but  can also be thought of as related to its $\beta$ function when the GW field's origin is the strength of a gauge coupling that spontaneously breaks conformal invariance  in the IR.  Here the $\phi^4$ term is consistent with conformal symmetry, which is weakly explicitly broken by the deformation of the marginal operator. This potential would generate a GW minimum at $k r_c =1/\epsilon \log{\frac{\lambda_1} {\lambda_2}}$.
 In a setup involving $M$ units of flux in the IR and a running coupling as in KS, we would  expect to equate $K/(M g_s)$ with ${1/\epsilon }\log{\frac{\lambda_1 }{ \lambda_2}}$.

In the case of KKLT, we don't expect a  radion lighter than the KK modes since  KK modes of the KS model (barring models with accidental symmetries and Goldstone bosons) have mass  set by the KK scale. If there were a light radion, we should see a light mode in the spectrum of KS in the field theory/decompactification limit. 
Numerical simulations as well as general considerations show that masses scale as the QCD string scale \cite{ks,numericalks, numericalks2, numericalks1} with no extra suppression.\footnote{ Note that Ref. \cite{numericalks} do find a light state, but that is associated with the spontaneous breaking of baryon number \cite{numericalks1}. We thank Matt Strassler for discussions on this point.} \footnote{ Note that other potential instabilities associated with the conformal factor were noted by \cite{Douglas} where it was shown they are tempered by the warp factor.}

We  now argue that we can explicitly identify such a radion in the full 10d theory. The above schematic radion potential is very close in form to the potential used in \cite{benarunning1} based on the analysis of \cite{dst, dt}. There they were describing the potential for the conifold deformation parameter defined in the warped region of the GKP geometry
in the local six-dimensional geometry corresponding to the deformed conifold by its embedding into \(\mathbb{C}^4\),
\begin{equation}
\sum_{a = 1}^4 \omega_a^4 = S \,.
\end{equation}
The deformation parameter $S$ is the complex structure modulus whose absolute value corresponds to the size of the 3-sphere at the tip of the cone.
\beq\label{eq:S}
\int_{A} \Omega_3= S \ , 
\eeq
The supersymmetric potential for this field induced by the Klebanov-Strassler geometry is
\begin{equation}
V_{KS} = \frac{\pi^{3/2}}{\kappa_{10}} \frac{g_s}{(Im\rho)^3}\left[c \log\frac{\Lambda_0^3}{\left|S\right|} + c'\frac{{g_s (\alpha' M)^2}}{\left|S\right|^{4/3}}\right]^{-1} \left|\frac{M}{2 \pi i} \log\frac{\Lambda_0^3}{S} + i \frac{K}{g_s} \right|^2  \,,
\end{equation}
where $g_s$ is the stabilized vev of the dilaton, $Im \rho=({\rm Vol}_6)^{3/2}$, 
\(c\) as we argue below is not relevant here (and is in any case suppressed in the small $S$ region), whereas the constant \(c'\), multiplying the term coming solely from the warp factor, denotes an order one coefficient, whose approximate numerical value was determined in \cite{dst} to be 
$c' \approx 1.18 $. 

We will now argue that essentially $S^{1/3}$ is the radion (up to normalization). The potential for the $S$ field is essentially the GW-induced radion potential presented earlier, but it takes a slightly different form than that above due to supersymmetry,
namely

\begin{equation}
    V=S^{4/3} \left(\lambda_1-\lambda_2 \log\frac{\Lambda_0^3}{S}\right)^2={\lambda_1^2}{S^{4/3} } \left(1-\frac{\lambda_2}{\lambda_1}\log{\frac{\Lambda_0^3}{S}}\right)^2\approx {\lambda_1^2}{S^{4/3}}\left(5- 6\left(\frac{S}{\Lambda_0^3 }\right)^{\epsilon}+2\left(\frac{S}{\Lambda_0^3 }\right)^{2\epsilon}\right)
\end{equation} 
where rewriting the potential in this GW form breaks down near the ``IR brane"  where $(2 \pi M/K) \log{S/\Lambda_0^3}$ gets big. Really the original form in terms of the logarithm is enough to see that we have weakly explicitly broken scale invariance. We only rewrite the term to show that schematically it agrees with the form of GW potentials in the literature.
Here $\lambda_1=K/g_s$ and $\lambda_2=(M/2\pi)$, and $\epsilon=\lambda_2/\lambda_1$. The minimum occurs at $S_{KS}=\Lambda_0^3 e^{-2 \pi K/M g_s}=\Lambda_0^3 e^{-\lambda_1/\lambda_2}=\Lambda_0^3 e^{-1/\epsilon}$. 
Here the $S^{4/3}$ dependence comes from the Kahler potential whereas the remaining dependence is from the superpotential.  The  nonrernormalization  theorems  in the supersymmetric potential guarantee the full potential is always proportional to the leading order potential. 

Notice that this form of the potential is consistent with an explicitly broken dilation symmetry, where $(M g_s)/K$ characterizes the breaking with $\lambda_2$ a spurion \cite{chacko}.  The scale invariance when $(g_s M)/K$ is small in combination with supersymmetry guarantee the above form of the potential, reinforcing the radion interpretation of $S^{1/3}$.

The above potential has both the minima we found before (the runaway $S=0$ as well as the desired GW minimum at $S=S_{KS}=\Lambda_0^3e^{-2\pi k/(Mg_s)}$ but also has  another unstable critical point, creating a barrier between the desired minimum and the runaway direction $S\approx\phi^3 \to 0$.

We can also expand about $S_{KS}$ to find
\begin{equation}
    V \approx S_{KS}^{4/3} \lambda_2^2 (\frac{S-S_{KS}}{S_{KS}})^2
\end{equation}
where this is the potential  when $S$ fluctuates away from the unperturbed minimum. We have not kept track of $M$-dependence here. The point is to see  that this potential corresponds to a superpotential for $S$  that is linear in the field near the minimum.  The nontrivial Kahler potential turns this superpotential into the mass term in the potential.   The GW potential takes this form in this supersymmetric setup. Notice that as expected the mass is $K$-independent and is not suppressed by $g_s M/K$.  When we use 
\begin{equation}
m^2_S \equiv \frac{1}{M_{pl}^2} G^{S \bar S} \partial_{\bar S} \partial_S V \Bigr|_{S = S_\mathrm{KS}} \,.
\end{equation} 
one finds \cite{benarunning1}  the $S$ mass squared is suppressed by $1/g_s M^2$. In terms of the properly normalized field $\phi$ (see below), the  mass squared scales (over the exponential suppression) as $1/(g_s M^2)^2 $, which is how all KK masses associated with the IR region of the conifold throat would scale as well.

From \cite{ks}, we see that the KK/glueball masses scale as $m^2/(g_sM)^2$ whereas the string tension scales as $m^2/(g_s M)$ so the ratio is $1/(g_s M)$ as it should be \cite{ks}. Here $m$ is a parameter from the metric defined in \cite{ks}, which is essentially $\phi/M$ to get the correct dependencies. Note that our formulas here, as in \cite{benarunning1}, are given in the Einstein frame.  Here $\phi$ is the properly normalized field related to $S$ by $\phi \propto S^{1/3}/\sqrt{g_s M^2}$. The string tension has the correct property that $1/g_s T_s^2\propto V_{antibrane}$, since $V_{antibrane}$ scales as $\phi^4/(g_s M^2)^3$ \cite{benarunning1}.

The   potential  above is a beautiful implementation of a GW type potential. The term from the Kahler potential that we kept here and that gives the nonperturbative radion mass derives from the radion kinetic term found by
 Douglas, Shelton and Torroba \cite{dst}, which   
 Bena, Dudas, Grana, and Luest \cite{benarunning1} as well as Blumenhagen, 
~Kläwer and Schlechter also presented.

 Ref. \cite{benarunning1} also identified the properly normalized field (including only the $c'$ term) in Eq. 3.16 as,\footnote{Ref. \cite{dt} also identified the physical field as $S^{1/3}$.} 
 \begin{equation}
\phi = \frac{3 M_{pl} \sqrt{c'}}{\pi^{1/2} \left\|\Omega\right\| V_w^{1/2}} \alpha' \sqrt{g_s} M S^{1/3} = \frac{3 \sqrt{g_s} M \sqrt{c'}}{8 \pi^4 \alpha' \left\|\Omega\right\|} S^{1/3} \,,
\end{equation}
where $c''\approx 1.75$.  Indeed the parameter $S$ is  related to $\phi^3$, and is the properly normalized field determining the warping in the throat.  This $\phi$ is precisely the radion of GW and has the correct potential to both determine the length of the throat and the warping in the IR as well as to respond to perturbations to generate a consistent geometry. 
The radion mass squared, as with the values of KK mass squared, is of size $1/(g_SM^2)^2\phi^2$, and is suppressed by a factor $1/(g_sM)$ compared to the string tension.

 The  interpretation of the $S$ field in the full 10d picture was the source of some confusion. The full kinetic term was given as \cite{benarunning1, dst}
\begin{equation}\label{eq:Smetric}
G_{S \bar S} = \frac{1}{\pi \left\|\Omega\right\|^2 V_{w}} \left(c \log\frac{\Lambda_0^3}{\left|S\right|} + c'\frac{{g_s (\alpha' M)^2}}{\left|S\right|^{4/3}}\right) \,.
\end{equation}
which behaved differently depending on the region of interest.  Ref. \cite{dt} distinguished the regime $r>\Lambda_0$, which would correspond to the bulk regime, $ (g_s M \alpha')^{1/2}<r<\Lambda_0$, which corresponds to the constant warp factor region, and the small $r$ region of interest, which is  the strongly deformed conifold region. The conifold deformation parameter  develops a kinetic term because a time-dependent fluctuation in $S$ modifies the 4d part of the metric, but not the warped 6d metric where the $S$-dependence cancels \cite{dt}. 

This cancellation is associated with the survival of the field $S$ in the decompactification limit, where the logarithmic $S$ dependence from the warp factor cancels that of the unwarped metric. 
The fluctuations in $S$  arise in the 4d theory and are controlled by the maximum redshift, which is exactly what we expect for a radion field. In \cite{kt}, this kinetic term, which gives the radion expected $S^{4/3}$-dependence,  was derived with a compensator formalism to allow kinetic terms for what from a higher-dimensional point of view seemed to be a 6d parameter. From a 4d perspective, this field is the radion that we knew had to be there in the stable throat picture and corresponds to the cutoff of the warp factor of the 4d metric in the full 10d picture.

The radion field we have identified is an infrared localized field that survives even when $M_{Pl4} \to \infty$.    The only region in which the $c$ term could have dominated is when $M_{Pl}$ is large due to large $\sigma$, which would allow for larger $\Lambda_0$, comparable to $M_{Pl4}$, the four-dimensional $M_{Pl}$. In other words, the the $c$ term depends on $\log{\sigma}$ but we know this dependence isn't part of the throat potential as it describes the dependence on $\sigma$ that comes from running the coupling only from the CY boundary and not from the string scale. Including this additional running should cancel that $\sigma$ dependence as above and leave only the $c'$ term in the region of interest, consistent with the radion potential we have found.

In \cite{benarunning1}, by keeping the kinetic term appropriate to the constant warp factor region dependent on $\log{S}$,  it was argued that the field is a modulus corresponding to a nonrenormalizable field in the decompactification limit. Such a  modulus should be a separate field associated with the gauge coupling that corresponds to the log piece of the kinetic term above but which is not part of the radion kinetic term. Ref. \cite{benarunning2} did nonetheless correctly observe that $S$ is the complex structure modulus whose absolute value corresponds to the size of the 3-sphere at the tip of the conifold, and furthermore that if one fixes a UV holographic cutoff, the running of the deformation parameter corresponds to changing the distance between the tip of the KS solution and the cutoff surface--again, what is expected for a radion. Moreover, when the throat is glued
into a flux compactification, they observed that this distance parameterizes the length of the so-called B-cycle,
which is fixed by the fluxes. These are the properties of the radion, but not of a modulus that decouples in the decompactification limit.

Having established the identify of the radion, we now have an exact formula for its potential which allows us to explicitly
evaluate any potential radion instability. Such an instability would be precisely the ``conifold instability" found in  Ref. \cite{benarunning2},  associated with the correctly identified radion field,  (and might be related to that of Ref. \cite{ buchel, mcg, benarunning1}).

So long as the perturbation is not too big, the radion  will adjust so that the GW field can redistribute the added energy. Adding the antibrane amounts to adding an energy source that scales as $\delta  S^{4/3}$ where $\delta$ is a small number. 

The antibrane contributes a perturbation
\begin{equation} \
V_{\overline{D3}} = \frac{ \pi^{1/2}}{\kappa_{10}}  \frac{1}{(Im\rho)^3}  \frac{2^{1/3}}{I(\tau)} \frac{\left|S\right|^{4/3}}{g_s (\alpha' M)^2} \,.
\end{equation}
We follow \cite{benarunning1} and define
$c'' = \frac{2^{1/3}}{I(0)} \approx 1.75 \,.$
For $p$ anti-D3 branes the potential is multiplied by $p$, and this is taken care by simply replacing $c''\to c''p$.

The general form of the potential (we factor out $\lambda_1^2 \pi g_s/c'$) is
\begin{equation}
    V=S^{4/3} \left(1+\epsilon \log{\frac{S}{\Lambda_0^3}}\right)^2+\delta S^{4/3} 
\end{equation}
The barrier disappears when $\delta/\epsilon^2=9/16$.

 However, one can readily see that this barrier disappears when $\delta/\epsilon^2=9/16$ which  is the instability found in Ref.\cite{benarunning1}. (Note that we are absorbing $g_s/K$ squared  in our definition of $\delta$ and $g_s/K$ in our definition of $\epsilon$  so that both are indeed small parameters.)  This would mean the field $\phi$ would no longer see the GW minimum, but would simply go onto the runaway direction; in other words the KS solution would have an instability.

We see that the perturbation from the antibrane (yielding the $\delta$ type perturbation above) yields the potential proportional to the above with  $\delta=  c''c' g_s/\pi K^2$ and $|\epsilon|=M g_s /2\pi K$. By writing it this way we keep $\epsilon$ and $\delta$ as small parameters. This gives precisely the stability condition found in  \cite{benarunning1}, namely
\begin{equation}
\sqrt{g_s} M > M_\mathrm{min} \,\qquad\text{with}\qquad M_\mathrm{min} = \frac{8}{3} \sqrt{ \pi c' c''} \approx 6.8 \sqrt{p}\,.
\end{equation} 
Although the scaling with $g_s M^2$ is as might be expected, this constraint is stronger by a numerical factor than one would anticipate and does indeed correspond to the instability of Ref. \cite{benarunning1}. 
If the above bound is not satisfied, the radion (conifold deformation parameter) will fall into a runaway direction if we perturb the original KS model with an antibrane. This instability is not the volume modulus instability. It is solely associated with the radion and would mean the ``IR brane" would go to infinity, or that the conifold deformation parameter goes to 0. This is possible because the binding energy (and the KK mass of the throat modes) is small, determined by the warped down QCD energy scale.  

Notice the potential we used when studying stability was the bulk potential. With a general GW potential, there can also be a brane contribution to the potential. It is not clear whether supersymmetry and the string construction could allow an additional term which could add  stability in principle. The only obviously allowed superpotential term that could affect a linear shift is proportional to $S T_{antibrane}$. Such a term would not necessarily help the stability criterion. It would allow for a shift in the constant term in the superpotential, but the effect would be to shift $S_c$ and change $\epsilon$ and $\delta$, but by an amount that is suppressed by $K$ and scales out of the stability criterion. The shift in flux itself from the antibrane, which reduces $K$ by one unit, would have an even bigger effect, but similarly leaves the stability criterion intact.

An antibrane is presumably not the only way to do  the uplift. Even if other physical theories in the IR are responsible for supersymmetry breaking, the radion stability bound is likely to apply. The energy in the IR will always be suppressed by $g_s M^2$, which is just a result of the warp factor,  relative to the contribution from the KS potential which will always be suppressed by $4 \pi^2$ so that presumably a lower bound on $g_s M^2$ comparable to the one we have found would still apply. This would include, for example, a low-energy theory that breaks supersymmetry based on gauge dynamics akin to that of the Standard Model, for example, in which case there would not necessarily be a spiky discontinuous landscape \cite{landscape1, landscape2, landscape3}. Advocates of the Swampland argue \cite{swampland} that such constructions are not possible within the constraints of a compactified manifold that stabilizes all the moduli. As we do not know all possible constructions, it is too soon to weigh in on this.  However, the challenge was stabilizing the modulus without generating AdS. Minkowski and de Sitter would be comparable and the stability bound above would still be relevant.

The analysis above (and in \cite{benarunning1}) was a perturbation to Minkowski space.   The additional contributions in perturbing AdS4 are further warp factor suppressed and are of size $V_{AdS4}S^{4/3}$, where $V_{AdS4}$ represents the energy associated with the gaugino condensate vacuum and as before $\phi \approx S^{1/3}$ where we have dropped factors of $M$.  An additional term would arise because a consistent geometry would require both the UV and IR boundaries to possess AdS4 geometry. The radion adjusts the bulk energy at the boundaries to provide consistent boundary conditions. 
We can also see this directly since we know the net energy would be that of AdS4 if the throat extended infinitely far (to the horizon of AdS5). Since there is a deformed conifold at nonzero $\phi$, the boundary has to carry the energy to make up the difference, which would be of order $V_{AdS4} \phi^4$. This follows since we can integrate the curvature from the UV to the IR brane, and would find the net curvature (without the additional boundary condition) would correspond to $V_{AdS4} (1-\phi^4)$. So there must be a  corresponding boundary term. Because $V_{AdS4}$ is already of order the warped down IR energy, this contribution would be doubly suppressed by the warp factor and can be neglected. This is in addition to a UV correction term, also essentially down by the warp factor (in magnitude) as well since it is a small correction to the leading UV boundary term for KS. Though this term has no $S$ dependence directly, it would also alter the stability criterion, albeit only by a warped down amount.

The instability  we  have identified would not be  a problem for $g_s M^2$ sufficiently large, consistent with the probe approximation for large but fixed $g_s M$, which is the regime in which the supergravity analysis applies. 
However, this bound has implications for the maximum size of the warp factor and also for cosmology evolution from high temperature. Before stating these, we mention two caveats and clarifications. First, we are assuming $g_sM$ is big by using the supergravity theory in the IR. If indeed $g_s M$ is truly not large, the supergravity approximation breaks down and the IR theory could behave very differently. Without explicit derivation of higher-order corrections, the boundary where the supergravity description breaks down is unclear. The bound we derive is an explicit numerical constraint.   Second, the constraint is strange in that the $\phi$ mass decreases as $g_sM^2$ increases, which appears to make the theory less stable. However, we can write the potential as
\begin{equation}
    V=m_{\phi}^2 \phi^2+\frac{1}{(g_s M^2)^3}  \tilde{V}_{antibrane}(\phi)
\end{equation}
where we rewrite the antibrane potential in terms of $\phi$ and have factored out $M$-dependence in $\tilde{V}$. The perturbation, which scales more strongly with the warp factor than the mass term (as $1/(g_s M^2)^3$ when written in terms of $\phi^4$ rather than $S^{4/3}$, decreases more quickly with $M$. Therefore even though the $\phi$ mass decreases with $M$, the stability to an antibrane perturbation increases. The constraint is on the perturbation relative to the original potential. Both decrease with large $M$ but the perturbation decreases more.

Returning to the instability, Ref. \cite{benarunning1}  argues  that for the real parameters one can get from known CYs it is difficult to both maintain a  reasonable hierarchy and not destabilize the system unless manifolds permitting larger net flux are found.   Their argument was that there is  insufficient wiggle room to get both large $K/(Mg_s)$ and large $M^2 g_s$ since $KM$ is bounded by the topology of the space.  For known stablized manifolds, the biggest hierarchy one can find in $\Lambda_{KS}/\Lambda_0$ is only about five once the stability criterion is imposed.  This is not ruling out KKLT in principle, particularly since large $g_s M$ is a part of the construction, but pointing out alternative geometric constructions are required to trust the result and stability of the system. A weakly bound state can be destabilized with only a small perturbation. In the event that suitable large flux manifolds are found, it is interesting that like the volume modulus mass, the glueball-sized radion mass   is  a consequence of quantum effects in the field theory.

In addition to the challenge of consistently generating a warped geometry and a KS geometry sufficiently stable to accommodate an antibrane, we note here another potential problem for the KKLT model with large $g_s M^2$. This one can't be solved by finding alternative CY geometries. Several authors\cite{cosmology} have noted the challenge to RS of cosmologically establishing an IR brane as the universe cools, a challenge that favors small values of $M$. The essential problem is that the high temperature universe would correspond to an AdS-Schwarschild black hole solution that must transition to the confining RS geometry at low temperature. Simple entropy arguments show that this transition is disfavored for large $M$ and explicit calculations show that the first order phase transition would be very suppressed, resulting in a supercooled inflating universe. Ref. \cite{jmr} worked out the implication of this constraint for a Klebanov-Tseytlin geometry. Their result is that the constraint on $M$ is strongest when the warping is small but even  for a warping sufficiently large to generate a  weak scale hierarchy,  the constraint as given \footnote{Possible ways to weaken the cosmological constraint have been considered \cite{recentcosmo}.} is $M^2<21$. This is clearly  inconsistent with the constraint $g_s M^2> 46$ derived above. So here again we find a challenge to the largish values of $M$ necessary to keep a stable KKLT geometry in order to generate a realistic scenario.

\section{ Supersymmetry Breaking}

Because it is independently of interest, we  will now ignore any potential instability and  consider the implications of the type of throat geometry we have introduced. We will show how the redistribution of energy is reflected in superymmetry breaking and also consider how supersymmetry breaking is communicated more generally.

We do the first part only schematically and leave the details of the full 5d supersymmetric warped geometry including the radion potential to later work (see \cite{sakai} for some steps in this direction).  Luty and Sundrum \cite{lutysundrum2} did a fully supersymmetric 5d unwarped model in which condensates in the bulk and on the branes stabilized the radion (their radion is more analogous to the volume modulus than to our radion)  and supersymmetry breaking occurred in the UV. As with the analysis above, one question  they addressed was how supersymmetry is communicated in the bulk space, in their case from the UV to the IR. The answer, as with any sequestered scenario, is anomaly mediation. This will be the case here as well, although for some fields the light volume modulus (which in some sense also gives an anomaly-mediated contribution as we will see) can give a comparable supersymmetry-breaking  contribution.

We will see that much of the communication of supersymmetry breaking can be understood from the chiral compensator through anomaly-mediation  although the volume modulus can play a role as well. However, the radion $F$ term is also important if we want to explicitly explore supersymmetry breaking communicated in the throat itself.

We distinguish  three different regions of the KKLT geometry. There is the QCD theory at the end of the throat, the throat itself, and the CY region (with possibly other throats). Because of the strong interactions at the end of the throat, supersymmetry breaking should be communicated to QCD scale KK modes more or less directly through operators involving only the strong interaction scale as argued in \cite{dewk}. In the absence of light fields, supersymmetry breaking would be communicated to the region outside the throat only through anomaly mediation, since supersymmetry breaking at the end of the throat is sequestered. The light fields we have considered are the volume modulus and the radion. Note that the  physical radion, though coupling in a similar manner to the chiral compensator,  is a distinct field,  and it too can communicate supersymmetry breaking. Ref. \cite{bht} also considered heavy gauge fields, which  have the potential to directly communicate supersymmetry breaking, but we will argue this is suppressed when the vector meson masses arise from supersymmetry preserving contributions in the UV.

Let us first explore how anomaly mediation should work in this case. We are interested both in the chiral compensator (useful for anomaly-mediation) as well as the radion (communicating supersymmetry breaking within the throat).  
Without warping, we expect $F_\chi\approx c/M_{pl}^2\approx m_{3/2}$ where $c$ is the square root of the energy in the supersymmetry breaking vacuum. With warping, we expect additional $\phi$ suppression. Notice that we have not explicitly included a supersymmetry breaking sector, which is presumably part of the QCD theory (since we know there is another vacuum that preserves supersymmetry). We are simply assuming the supersymmetry breaking energy which requires a nonvanishing chiral compensator, with energy corresponding to the supersymmetry breaking.

Again, we are not doing the full warped supersymmetric Lagrangian so we present a schematic to show how we anticipate supersymmetry breaking will be communicated. We expect $F_\chi \approx m_{3/2} \approx \phi^2 \sqrt{\delta T}/M_{Pl}$ since only the additional antibrane energy breaks supersymmetry and its contribution should be warped.     The dependence on $\phi$ should come from the  supersymmetry-breaking superpotential which should include a factor of  $\phi^3$ when we use the properly normalized field $\phi$ but from the Kahler potential when we worked in terms of $S$.

To get further insight into the uplift, it is also of interest to better understand the communication of supersymmetry breaking inside the throat, where direct interactions can in principle give important contributions. The radion, in particular, should develop a supersymmetry-breaking $F$ term that could in principle allow the ``GW field" to  communicate supersymmetry breaking to radion-dependent operators.
We now argue that the supersymmetry breaking uplift energy is located throughout the bulk so that it is consistent with the geometry described above. 

 From a 4d perspective, we know the IR brane net energy should be the warped down version of the net UV brane energy (with opposite sign), which is assumed to be be  of order the small cosmological constant today. The initial energy on the UV brane was the initial AdS energy from the gaugino condensate whereas the initial energy on the IR brane is the uplift energy (of the same order in absolute magnitude as the gaugino condensate energy).  We showed earlier how the junction conditions of the GW field allow the extrinsic curvature to compensate for the mismatched energy, allowing for slicing with a small de Sitter constant throughout the space, and in particular on  both the IR and the UV boundaries. For this to happen, the net supersymmetry breaking energy will be almost zero in the IR, since the energy on the IR brane, which is associated both with direct supersymetry breaking through the antibrane and the extrinsic curvature from the IR matching,   has to add to the warped down de Sitter energy, which is  essentially zero.  On the other hand, the supersymmetry breaking contribution should be of order the uplift energy on the UV brane. That is, although supersymmetry breaking happens in the IR, the net effect of supersymmetry breaking has to be that the uplift energy is essentially all on the UV brane (or in the boundary matching there) (with warped down energy throughout the rest of the space). The only field in our effective theory that can allow this to happen is the radion, which must develop an $F$ term that allows the GW field to cancel energy in the IR and add it in the UV.
 Note that aside from the boundary matching, we can generally  ignore the radion when considering supersymmetry breaking. However, we have seen above that the radion is critical to consistency of the higher-dimensional theory and furthermore is necessary for studying stability.

 In the 5d theory, what distinguishes the radion potential is that the leading contribution arises from the matching associated with the kinetic term for $\Phi$.  We expect the matching to be similar to that of the GW analysis above since the $D$ term $\left (Z(\mu/\mu_0) \Phi^\dagger \Phi\right)_D$ is essentially the kinetic term.  In both cases, (up to a factor of two) one derivative (or equivalently $\int d^2 \theta$) is  associated with the initial field derivative while the other spatial derivative is replaced by the change in the field which is here accounted for by the $F$ term of the radion.   In the 4d theory, we have only scale dependence (rather than $r$ dependence) but dependence on the scale $\mu_0$ should translate into $\phi_c$ dependence in the 5d theory. 
We expect the derivative wrt $r_c$ of the GW field to correspond to the derivative wrt $\mu_0$ of  $Z(\mu_0)$ which should give us $\gamma$. Here  $\gamma$ (rather than a scaling directly of the superpotential) can be thought of as the source of the RG running (and is explicitly in \cite{gwkklt} for example. Solving for the $F$ term in terms of the IR boundary condition with the correct solution to the equations of motion in the 5d theory should be equivalent to the matching analysis we did above.
This would require that the radion develops an $F$ term that scales with the brane energy, and furthermore is proportional to $1/\gamma$ which  in the IR is determined solely by $M$.

If there is a supersymmetry breaking perturbation, $\delta T$, the above matching would require an $F$ term for the radion,   to  be roughly of the form
$F_\phi/\phi\approx \delta T /\gamma $ where $\gamma$ from the bulk potential would scale as $(M/2\pi)$.  This term is interesting in that it is proportional to the SUSY breaking energy (not its square root).

With the explicit supersymmetric radion/conifold-derformation-parameter, we can investigate  how supersymmetry breaking is transmitted through the radion field.
The superpotential for the radion in the supersymmetric theory is of the form $1+ \gamma \log {\phi}$, where the second term plays the role of $\phi^\gamma$ in the usual GW potential. It is precisely this term that allows the radion to shift in the way that is required and is exactly the type of shift we were exploring when studying the instability. 
We expect supersymmetry to be broken in the QCD sector leading to an effective term in the superpotential $ \sqrt{T_{antibrane}}\phi^2 \chi$ where $\chi$ is the chiral compensator or alternatively can be done in terms of a nilpotent goldstino field. Such a term would add the necessary uplift energy.  This will give the uplift and the stability criterion we explored earlier such that the contribution to the potential from $F_\chi^2$ yields $T_{antibrane}$, the uplift we included when studying the full radion potential.

We can take a step further by studying how the radion shifts for small uplift energy. If we define $X=\epsilon \log{(S/\Lambda_0^3)}+1$, so that $W \supset S K  X$, we find in the presence of a perturbation $\delta$ (which in KKLT had energy density proportional to $T_{antibrane} S^{4/3}$)
\begin{equation}
    X=-\frac{3}{4}\epsilon \pm \sqrt{\left(\frac{9}{16} \right)\epsilon^2-\delta}
\end{equation}
which tells us that 
\begin{equation}
    F_S\approx K  \Delta X\approx- K \frac{\delta}{\epsilon}= -\frac{T_{antibrane}}{M} 
\end{equation}
which is the form of the shift required for consistency of the construction.

We also see that although the $\phi$ mass is comparable to other KK masses, the $\phi$ term is the one that develops this VEV due to the $\phi$-dependence of the antibrane potential--both this term and the supersymmetry breaking term above. Moreover because of sequestering which means the kinetic terms for the UV and IR fields do not mix, we do not expect any significant direct interactions between the gaugino condensate and the antibrane.

Since $F_S$ will not play a role aside from guaranteeing internal constency in the throat, which we know should apply so long as the theory is not destabilized, we move on to more general communication of supersymmetry breaking. From the perspective of  fields located away from the throat, supersymmetry breaking is sequestered so supersymmetry-breaking will be communicated through anomaly-mediation. In this case the  communication of supersymmetry breaking can be seen as a consequence of the $F$ term of the  conformal compensator \cite{rs0}. Such sequestered supersymmetry breaking will be reflected in the dependence of any operator coefficient on the scale factor and supersymmetry breaking occurs through violations of conformal invariance.  We can derive these anomaly-mediated contributions  to the nonthroat fields by going to the 4d theory. Note that in the 4d effective theory, terms can also be suppressed also by the warp factor appropriate to a field's location in the higher-dimensional space.

Anomaly mediation applies in the absence of other light fields. Because the volume modulus is light and  there is  $\sigma$-dependence in the supersymmetry breaking antibrane term,  the volume modulus can also get a supersymmetry-breaking $F$ term in which case there is ``mixed modulus anomaly mediation". \cite{cfno}.  
 
 This would lead to $\sigma$ developing an $F$ term  $F_\sigma/\sigma \approx \frac{F_\chi} {a \sigma}$, which is parametrically suppressed by $a \sigma$. We can explicitly derive this result  from the term in the supersymmetric Lagrangian \cite{cfno} $e^K K^{ij} (D_{\sigma} W)^\dagger (D_{\sigma} W)$, where in the presence of the perturbation we would no longer find $D_{\sigma} W=0$
 
 Note that supersymmetry breaking occurs in the QCD sector \cite{dewk} so there can be nonvanishing $F$ terms there that communicate supersymmetry breaking as well, but that will be only to other fields at the end of the throat. Assuming no other light fields aside from the KK modes at the end of the throat, only the conformal compensator $\chi$ and the volume modulus $\sigma$  communicate supersymmetry breaking to the rest of the space. Ref. \cite{bht} argued that if there are vector fields under which both UV and IR supersymmetry breaking fields transform, there can also be $D$ term contributions. However, $D$ terms don't contribute at leading order in $1/M_V^2$ if $M_V$ has a supersymmetric origin, as it does for the UV contributions in this case \cite{Dterm}. The contributions only arise with supersymmetry breaking so have additional suppression by the SUSY breaking scale divided by the vector mass squared. This is seen both from nonrenormalization theorems and a cancellation in superspace between a direct four-point function and vector exchange in the supersymmetric limit.

 If $a \sigma< 8 \pi^2$,   the modulus-mediated contribution would  dominate over anomaly-mediation  in $\sigma$-dependent terms (which is implicitly assumed in Ref. \cite{kt}). Ref. \cite{cfno} argues that this inequality is always satisfied or at best saturated for a KKLT construction and work in parameter space where they argue modulus and anomaly mediation give comparable contributions. We leave this an open question.
 
 Notice however that if we think of $a$ as the inverse of the QCD $\beta$ function (more precisely the beta function/3  divided by the squared coupling), 
 the modulus-dependent term is also of the form of an anomaly-mediated contribution, suppressed by the beta function and by $g^2/8 \pi^2$, which by assumption is determined by $1/\sigma$.  \footnote{In fact, if the $\sigma$-dependence of the antibrane was $\frac{D} {\sigma^3}$ we would get precisely the beta function. Presumably the additional scaling of the antibrane as it runs down the throat changes this scale dependence.} \footnote{This answer is  different from that in \cite{lutysundrum2}, where the $F$ term for the ``radion" aka volume modulus actually grew with the size of the space. The difference is that in the Luty-Sundrum supersymmetry breaking model, the ``radion" field played a role in supersymmetry breaking. If gravity decoupled, there would still be global supersymmetry breaking due to the radion $F$ term.} Whether this term contributes to supersymmetry breaking depends on the $\sigma$ dependence of the relevant fields and parameters, but it is interesting that it is contributing comparably as if it were an anomaly-mediated contribution as well.

We can also see that the volume modulus supersymmetry breaking tadpole is linear in $F_\chi$, whereas the radion supersymmetry breaking tadpole is quadratic--that is, it is directly proportional to the perturbation and not its square root.   
The radion tadpole, as we found earlier, is determined by  the supersymmetry breaking perturbation to the IR brane energy.

  Our mass scales on the whole agree with those in Ref. \cite{kt}.  However, we do note that there are additional light fields (compared to the CY KK modes and string scale modes), including the radion and other KK modes associated with the IR condensate  at the end of the throat. Since they have mass squared of order $\phi^2/((g_s M^2)^2\sigma^{3/2})$  and the antibrane tension scales as $\phi^4/((g_sM^2)^3)\sigma^3)$, which is approximately equal to the gaugino  energy which is $m_{3/2}^2 M_{pl}^2$ , we see that  these modes should have mass $m_{3/2} \frac {M_{Pl}}{ M_5}/ \sqrt{{g_S} M^2 }$.   Their interactions should be small away from the end of the throat but as argued above, the radion plays a role within the throat.
  
  In the following section, we  consider some interesting implications of this form of sequestering.
We neglect any direct moduli-dependent supersymmetry breaking  and focus on anomaly-mediated supersymmetry breaking communicated by the chiral compensator.

  \subsection{Zoned Anomaly Mediation}

  The combination of warping and sequestering is very natural in this type of set up. Therefore anomaly-mediated supersymmetry-breaking should be expected.   Theories with a similar setup were considered in \cite{ramankklt,gp}. In those theories supersymmetry breaking was communicated more directly to a fundamental sector of the theory in the UV, but some masses for composite states in the IR, well-isolated from UV supersymmetry breaking, occurred through anomaly mediation. These theories all have in common the absence (or suppression) of bulk scalar mass terms, which scale to zero in the IR, an  observation made also in \cite{strassler4,kt4}, in which  supersymmetry breaking effects are naturally set only by the IR scale. 
  
  It is of interest that in such extensions of the type of construction we have considered there can  potentially be interesting hierarchies of masses originating in different throats, which could also have important implications for moduli interactions.   Despite a common Planck scale in the 4d theory, different regions of space can experience supersymmetry-breaking very differently, depending on where they are located and the metric or warping in that region of space, in what we call ``zoned anomaly mediation".

    In this case, fields will experience anomaly-mediation, but there can be warp factor suppressions on top of the usual loop factor. In anomaly-mediated setups,  superfields generally obtain supersymmetry breaking masses proportional to $\alpha m_{3/2}$, where $\alpha$ is associated with the assumed nonzero running. However,  a throat is a consequence of approximate scale invariance so masses in throats would generally be further protected \cite{strassler4, kt4, gp, ramankklt}.  This might be the result of  a global symmetry or supersymmetry protecting approximate conformal  invariance, in which case the only mass
  scale from the throat would come from the IR. 
  
  The important point is that so long as the supersymmetry breaking is restricted to the strong gauge sector generating the throat and enough symmetry is preserved, anomaly-mediated supersymmetry breaking masses will appear only in terms determined by the IR scale, even if fundamentally the breaking was in the UV. 
  In  \cite{strassler4}, for example, supersymmetry is badly broken in the UV. Nonetheless, so long as sufficient global symmetry is preserved, the theory then implicitly ``knows" it will run into the IR so the conformal warp factor in the IR would give suppressed supersymmetry breaking to leave the UV theory under perturbative control.  This could be true for example in  Ref \cite{strassler4}. Even if supersymmetry were entirely broken in the throat, it is not necessarily badly broken in the rest of the theory.

   We expect that supersymmetry breaking masses, like all other masses for fields not localized in the UV, will be warped (and potentially volume suppressed) so that they will run to small values determined by the IR scale. Unless anomaly mediation breaks the global symmetry that protected the throat, the throat will survive. Since anomaly mediation communicates only $R$ symmetry breaking, any nonabelian symmetry or chiral symmetry would survive (for example SO(6) in \cite{strassler4}). If there were no gauge single relevant operators before supersymmetry breaking, there still would not be any after. Gauge singlet marginal operators can exist, but we expect them to still be associated with an IR scale.

  In detail,  fermion masses that would have destroyed the throat were already forbidden by global symmetries and scalar masses that could have been dangerous correspond to relevant operators that scale to a small value. In the IR, where conformal invariance is badly broken,  masses are no longer protected. The strong bound states are in supersymmetric multiplets, but experience anomaly mediation, which is generally maximal since the gauge coupling is strong. 
  
  There can also be fundamental fields (that is fields with wavefunction not localized in the IR). These can include heavy matter particles or gauge bosons and their partners. Indeed fundamental matter localized in the UV can experience unsuppressed SUSY-breaking masses, but these splittings will be sequestered from the fields situated in the IR. Gauge bosons can in principle be a problem in that they extend from the UV to the IR. This introduces two potential dangers. First is SUSY-breaking $D$ terms and the second is that the gauge sector, which does experience mass splitting in the UV, couples also to fields in the IR, apparently violating the sequestering assumption. The first issue is addressed when there are no gauged U(1)s. The second issue is addressed when, as above, there are no global singlet relevant operators. Explicitly, \cite{ramankklt} shows that so long as the spectrum of the N=2 representation of supersymmetry arising from the 5d superymmetric theory is maintained, the only fields that are lifted by supersymmetry breaking are the original gaugino and the original adjoint scalar in a multiplet with the remaining adjoint fermion. These fields can pair up to make an approximately supersymmetric gauge multiplet so that  mass splittings in the composite sector remain suppressed by the IR scale.

    So even within the framework of anomaly mediation, there can be hierarchies of masses, over the usual ratios of couplings because supersymmetry breaking can be even further insulated from sectors of the theory that correspond to a throat than in conventional scenarios of anomaly mediation.   This can have important phenomenological implications, since sequestered sectors of the theory can involve apparently very different susy-breaking scales. Moreover, if we live in a warped throat, the gravitino mass can be much heavier than naively anticipated if supersymmetry is broken in another region of the space.  Of course within any interacting sector we do not expect such hierarchies as noted in \cite{split3}.  In particular, all the particles of our sector that interact with each other presumably come from the same sector. Only hidden sectors might have unusual hierarchies of masses.

  Notice we also have the possibility to break supersymmetry multiply in different throats.  From the 4d point of view, the theory is just anomaly mediation but the ``radions" of the different sectors are stabilized by the field theory, independent of 4d gravity. This is simply the statement that we can consistently have different gaugino condensate scales that break supersymmetry independently. Since the theories are sequestered, based on locality we do not expect Kahler potentials to connect the different sectors, which are connected only through gravity.  Supersymmetry breaking can be broken dynamically by different sectors where the gaugino condensate scale corresponds to different warp factors as well as different overall volumes.  In this case, the biggest contribution will dominate anomaly mediation, but in warped regions, will convey an apparently warped down scale of supersymmetry breaking. The gravitino mass will primarily be determined by the largest supersymmetry breaking term. However the dominant goldstino interactions in each zone can be different.

 We also note that sequestered  light particles might be more   stable than expected from a 4d perspective. The scaling to get the 4d $M_{Pl}$ is inessential to the field theoretical stabilization in the throat, where it can essentially act as global supersymmetry breaking if supersymmetry breaking occurs in that throat, for example. Any higher energy from outside the throat will get warped down  so it won't necessarily destabilize anything. This will depend on time scales and whether the different regions reach thermal equilibrium. The differerent CFTs corresponding to different throats would be expected to interact only through Planck-suppressed operators so might never reach thermal equilibrium with each other and in fact have quite weak interactions with each other, often more than Planck-suppressed  \cite{benedict}. This necessity for accounting for the metric when determining stability might also be true in other situations with disconnected light states, such as  the Swampland. \cite{largefieldswampland, swampland}.

\section{Conclusions}
In this paper we have argued that some of the critical issues with KKLT can be understood from a purely five-dimensional gravitational perspective. The requirements on the five-dimensional geometry are inconsistent with a static solution unless the radion and its back-reaction on a GW-like field is explicitly included.  
 Some of these issues are irrelevant to the 4d effective theory. However, the radion needs to be included in this theory to check for stability. Most papers supporting KKLT and those objecting were focusing on the issues of stability of the antibrane and the potential for destabilizing the volume modulus and neglected the existence of a relatively light radion in the throat. One interesting exception is \cite{benarunning1}, which suggest a ``conifold instability" which is indeed a radion instability that would occur if $g_s M^2$ is too small. We have argued that the field associated with the conifold deformation parameter is a radion from the perspective of the 5d effective theory.  The requirement of larg(ish) $g_sM^2$ is not  surprising given that we already knew that $g_s M$ and $M$ should be large for the approximations to be trustworthy. What is new is a precise numerical target where we learn that $g_s M^2> 46$ for the construction to be stable, which is perhaps slightly bigger than naively expected though consistent with strong coupling involving $2 \pi$. This constraint forces constructions allowing larger fluxes than might have been expected, potentially introducing additional moduli fields and further challenges to a fully stabilized theory.  We have argued that even with alternative sources of an uplift, such a constraint is likely to survive and might in any case be incompatible with a vialbe cosmology.

We have also considered more generally the communication of supersymmetry within this type of construction and have identified the natural connection between  warping and sequestering and argued that anomaly-mediation should play the dominant role in most communication of supersymmetry breaking in this type of setup. Potentially, in generalizations,  different sectors of the theory can generate very different anomaly-mediated masses.  It will be very interesting to pursue both this and possible KKLT generalizations in the future.

\acknowledgments
Many people were critically important to the development of the ideas in this paper. Prateek Agrawal, Michael Douglas,   Severin Luest,  and Matt Strassler in particular were all very generous with their expertise but I also benefited tremendously from conversations with Nima Arkani-Hamed, Iosif Bena, Csaba Csaki, Ulf Danielsson,  Thomas Dumitrescu, Emilian Dudas (who shared his recent related work), Dan Harlow, Andreas Karch,   Curt McMullen,  Lubos Motl, Alex Pomarol, Matthew Reece, Savdeep Sethi, Cumrun Vafa, Irene Valenzuela,  Benedict Van Harling, Zhong-Zhi Xianyu, and Max Zimet. Part of this work, particularly Appendix A,  originated with David Pinner and I am grateful for his participation.  This research is supported by NSF
grants PHY-1620806 and PHY-1915071, the Chau Foundation HS Chau postdoc support
award, the Kavli Foundation grant “Kavli Dream Team”, and the Moore Foundation Award 8342.

{\appendix} \section{10D Einstein Equations}  \label{AppendixA}
Although tangential to the main part of the paper, it is interesting to see the challenges that arise in trying to obtain de Sitter space from a compactified higher-dimensional space purely from a gravitational perspective.\footnote{We thank David Pinner for collaboration on  the work presented here.}
For an effective theorist, getting a dS vacuum  might seem trivial in that you can write down the 4d theory you want.  However, we require the existence of a UV higher dimensional theory that justifies the model and therein lies the challenge.

 We will consider several simple cases that reduce to 4D de Sitter, or to 5D  spaces sliced by 4D de Sitter (since  all  we need to find is vacua that {\it look} like de Sitter vacua to a four-dimensional observer). Here, unlike in critical string theory, we allow for a cosmological constant. 
Nonetheless, in  agreement with previous No-Go Theorems \cite{mn,underwood}, we find  that getting de Sitter vacua  appears to require  hard-to-realize energy-momentum tensors, at least in simple warped or factorizable geometries which are in principle the most straightforward places to look.  We find perhaps the most natural possibility (with our restrictive assumptions) turns out to be RS de Sitter in five dimensions. The challenge as with KKLT would be to get appropriate boundary conditions with a fully stabilized moduli space.

We first assume a factorizable geometry and ask what it would take to get an energy momentum tensor that would allow de Sitter space in four dimensions with the other additional compactified dimensions static.  We consider Einstein Equations only and don't impose supersymmetry or string constraints.  

First consider a metric which is de Sitter, with the extra dimensions curled into a flat torus with constant radii.
\begin{equation}
    ds^2 = -dt^2 + a(t)^2 \delta_{ij} dx^i dx^j + b^2 \delta_{AB} dx^A dx^B
\end{equation}
Here the indices $i,j$ run over $[1,..,3]$ and $A,B$ run over $[4,..,n]$.  Defining $H \equiv \dot{a}/a$, this corresponds to a stress-energy tensor of the form
\begin{equation}
    T^M_N \propto -3 H^2
    \begin{pmatrix}
    1 &&&& \\
& 1 + \frac{2}{3} \frac{\dot{H}}{H^2} &&& \\
&& \ddots && \\
&&& 2 + \frac{\dot{H}}{H^2} & \\
&&&& \ddots
    \end{pmatrix}.
\end{equation}
Note that this stress-energy tensor is not proportional to the metric, even when $\dot{H} = 0$.  Some additional contribution is required--one that is not realized in the energy-momentum for the usual types of sources considered.  In the presence of a positive $n-$dimensional CC with $8 \pi G \Lambda = 3 H^2$, additional anisotropic pressure would be required in the form $\delta T^A_B = \delta p\ \delta^A_B$.  Here, again, $A,B$ run over $[4,..,n]$.  Such a stress-energy contribution seems odd, since there is no corresponding energy density, $\delta T^0_0 = 0$.  
\footnote {Ref. \cite{adk} also considered the possibility of de Sitter with constant extra dimensions. The solution they found when considering only the lower-dimensional Einstein Equations corresponds to a tuned flat direction in the higher-dimensional space and would not give rise to consistent cosmology with the additional dimensions remaining stable. }

Alternatively, a positive $n-$dimensional CC equal to $8\pi G \Lambda = 6 H^2$ could be paired with an additional component of the form
\begin{equation}
    \delta T^M_N \propto 3 H^2 
    \begin{pmatrix}
    1 &&&&& \\
& 1 &&&& \\
&& 1 &&& \\
&&& 1 && \\
&&&& 0 & \\
&&&&& \ddots
    \end{pmatrix}.
\end{equation}
From the perspective of the extra dimensions, this fluid component appears to behave like matter, while it behaves like vacuum energy in the other four dimensions.  One interpretation is that this corresponds to a gas of D3 branes filling the $n-$dimensional space.  However, the sign of $\delta T$ is opposite to that of $T$: these branes have negative tension and would need to be stabilized.
A stable realization of this scenario could be a D3 brane at an orbifold fixed point or an orientifold with the extra dimensions sufficiently small so that we can average the D3 brane energy to effectively get a negative 4d cc spread throughout the space (even though in reality concentrated at an approximate delta function). Here we would presumably need an additional stabilization mechanism to realize such a scenario consistently.

Another potentially more promising approach is to include a curvature degree of freedom in the additional dimensions. This is a natural possibility suggested by the energy momentum tensor in that it can consistently change only spatial components of the energy momentum tensor, allowing for the desired freedom to get the energy momentum tensor that is required. 
We do this explicitly for a case that is a  simple but unstable example in which the extra dimensions are curled into a sphere, using the curvature to balance the expansion of the bulk CC.  For simplicity, let's consider $dS_4 \times S_2$.  Again, we'll take both $H$ and the radius of the sphere ($r$) to be a constant:
\begin{equation}
ds^2 = -dt^2 + a_0^2  e^{2 H t} \delta_{ij} dx^i dx^j + r^2 \left( d\theta^2 + \sin^2\theta d\phi^2\right).
\end{equation}
This corresponds to the following stress-energy tensor,
\begin{equation}
    T^M_N \propto -\frac{1}{r^2}
    \begin{pmatrix}
    1 + 3 H^2 r^2 &&& \\
& \ddots && \\
&& 6 H^2 r^2 & \\
&&& 6 H^2 r^2 \\
    \end{pmatrix},
\end{equation}
so that this metric is consistent with a positive bulk CC so long as $r^{-2} = 3 H^2$.  However, fluctuations around this solution ($a(t) \rightarrow a_0 e^{H t} + \delta a(t)$, $r \rightarrow (\sqrt{3} H)^{-1} + \delta r(t)$) are unstable, with $\delta r(t)$ growing exponentially.

In summary, as with Refs. \cite {mn, underwood, gibbons},  we find the required form for the energy momentum tensor  is unlikely to be realized by stable physical objects (though there is a solution with bulk AdS and a negative tension brane). However, we see that if there is freedom to include arbitrary curvature from the extra dimensions, a consistent solution can be found.

Maldacena and Nunez and others considered a  more general warped case. Since all one wants is a low energy four-dimensional theory  with a small positive cosmological constant, we can in fact compactify to five-dimensional space so long as the boundary conditions induce 4d de Sitter slicing. We will first show this in a theory with five dimensions and then extend the analysis to higher dimensions. 

Although we allow general warping, we will find that the only solution with the above assumptions has a negative cosmological constant, but one which has $dS_4$ slicing so that it looks like expanding de Sitter space from a four-dimensional perspective. This is exactly the warped solution sliced by de Sitter space found in refs \cite {kr, dfk, cvetic1, cvetic2, kaloper, kimsquared, nihei},

To show this, we allow for general warping with respect to the extra dimension, with the restriction that 4d slices have a metric that looks like de Sitter space.     In 5d, we're left with the following effective line element:
\begin{equation}
    ds^2 = -f(y)^2 dt^2 + g(y)^2 e^{2 H(y) t} \delta_{ij} dx^i dx^j + dy^2.
\end{equation}
In order that the stress-energy tensor corresponding to this metric has no off-diagonal components, we must have
\begin{equation}
    H(y) = H,\hspace{1cm} g(y) \propto f(y),
\end{equation}
so that
\begin{equation}
    ds^2 = f(y)^2 \left(-dt^2 + a_0^2 e^{2 H t} \delta_{ij} dx^i dx^j\right) + dy^2
\end{equation}
and
\begin{align}
G^{\mu}_{\nu} &= -\frac{3}{f(y)^2} \left[H^2 - \left(f'(y)\right)^2 - f(y) f''(y) \right] \delta^{\mu}_{\nu}, \\
G^5_5 &= -\frac{6}{f(y)^2} \left[H^2 - \left(f'(y)\right)^2\right],
\end{align}
in which the Greek indices run over four dimensions.  This Einstein tensor will be proportional to $\delta^M_N$ so long as
\begin{equation}
    f(y) f''(y) = -H^2 + (f'(y))^2.
\end{equation}
Taking an exponential ansatz for the solution, $f(y) = A \exp (-k y) - B \exp (k y)$, the equation above is satisfied for any choice of $A$ and $B$ with $H^2 = 4 A B k^2$.  For real $H$ and $k$, $A$ and $B$ must have the same sign.  Despite the apparent $dS_4$ slicing, this corresponds to a negative bulk CC:
\begin{equation}
G^M_N \rightarrow (\frac{3 H^2}{2 A B} )=-k^2 \delta^M_N
\end{equation}
The Einstein field equations, $M_*^3 G^M_N + \Lambda g^M_N$ = 0, relate the constants to the CC; for $H$ real and positive, we must have $\Lambda < 0$. The warp function, then, contains both the usual exponentially decaying solution as well as an additional exponentially growing solution proportional to the $dS_4$ Hubble constant,
\begin{equation}
f(y) = A e^{-\sqrt{\frac{|\Lambda|}{6 M_*^3}} y} - \frac{3 H^2 M_*^3}{2 A |\Lambda|} e^{\sqrt{\frac{|\Lambda|}{6 M_*^3}} y}
\end{equation}

This is precisely of the form of a bulk AdS metric with curvature $k=1/L$ with de  Sitter slicing. To complete the analysis in this case and realize a potentially physical geometry we would need to impose boundary conditions. As advertised, this could lead to AdS de Sitter. 

There is an important interesting general lesson here. Space can appear to be de Sitter from a four-dimensional perspective, even though the overall cosmological constant and even that for a lower-dimensional five-dimensional world is negative--in principle  more accommodating to supersymmetry at higher energies.

Now we'll see how the solution is modified if we enlarge to a product space. we consider here a simple attempt with an extra compact sphere.
\begin{equation}
ds_{10}^2 = f(y)^2 ds_4^2 + dy^2 + r^2 d\Omega_5^2,
\end{equation}
in which $ds_4^2$ is the 4-dimensional de Sitter line element, as above.  The Einstein tensor is modified:
\begin{align}
G^{\mu}_{\nu} &= -\left(\frac{3}{f(y)^2} \left[H^2 - \left(f'(y)\right)^2 - f(y) f''(y) \right] + \frac{10}{r^2}\right) \delta^{\mu}_{\nu}, \\
G^5_5 &= -\left(\frac{6}{f(y)^2} \left[H^2 - \left(f'(y)\right)^2\right] + \frac{10}{r^2}\right), \\
G^A_B &= -\left(\frac{2}{f(y)^2} \left[3 H^2 - 3 \left(f'(y)\right)^2 - 2 f(y) f''(y) \right] + \frac{6}{r^2}\right) \delta^A_B,
\end{align}
where, as before, the Greek $\mu$, $\nu$ indices run over four dimensions and the capital Latin indices now run over the spherical coordinates.  Requiring that the 4d elements of the Einstein tensor are equal to the 5th diagonal component gives the same condition as above,
\begin{equation}
f(y) f''(y) - \left(f'(y)\right)^2 + H^2 = 0.
\end{equation}
Plugging in the same ansatz as above leads to an Einstein tensor that, while independent of $y$, is not proportional to the identity for all values of $r$:
\begin{align}
G^{\mu}_{\nu} &= \left(\frac{3 H^2}{2 A B} - \frac{10}{r^2} \right)\delta^{\mu}_{\nu}, \\
G^5_5 &= \frac{3 H^2}{2 A B} - \frac{10}{r^2}, \\
G^A_B &= \left(\frac{5 H^2}{2 A B} - \frac{6}{r^2}\right) \delta^A_B.
\end{align}
This geometry will be sourced by a bulk CC only when
\begin{equation}
r^2 = -\frac{4 A B}{H^2}=-k^2
\end{equation}

As described above, real and positive $H$ requires $A$ and $B$ to have the same sign, so that this condition amounts to negative curvature in the extra dimensions with curvature set by the AdS scale of the 4D space.  (Note that there is another solution for which $k$ is imaginary, $A$ and $B$ have opposite signs, and both $H$ and $r$ are real and positive, but it is sourced by a positive bulk CC rather than a negative one.) This has some overlap with what was found for string constructions, where a negative curvature ingredient can play an essential role \cite{evaneg}.

\section{5d Cosmology}
We review here how cosmology works in the presence of a stabilizing field as was done in \cite{cgrt}.
In terms of a metric
\begin{equation} 
ds^2=n(y,t)^2 dt^2-a(y,t)^2 (dx_1^2+dx_2^2+dx_3^2)-b(y,t)^2 dy^2,
\nonumber \\
\equiv \tilde{g}_{AB}(x,y) dx^A dx^B.
\end{equation}
with two branes are located at $y=0$ and at $y=1/2$ 
Ref. \cite{cgrt} parameterized
 the equation for the perturbation in the metric using the ansatz
\beq a(t,y)=e^{Ht} e^{-|y|m_0b_0} (1+\delta a(y)), \ \ \
n(t,y)=e^{-|y|m_0b_0} (1+\delta a(y)), \ \ \ b=b_0.
\eeq

With the general leading order solution
\beq
\delta a(y)= \frac{\alpha}{4m_0 b_0} 
(e^{4 |y|b_0 m_0}-1)-\frac{H^2}{4 m_0 ^2} (e^{2 |y|b_0m_0}
-1),
\label{da}
\eeq
where the overall constant has been fixed such that $\delta a(0)=0$.
The remaining two constants, $\alpha$ and $H^2$ have to be fixed such 
that the jumps of this function at the two branes reproduce the 
matter perturbation that we are including. The result is given by
\beq
H^2=\frac{\kappa^2 m_0}{3 (h1-\Omega^2_0)} (\delta V_* +\delta V \Omega^4_0),
\label{hh}
\eeq
and the value of the other constant $\alpha$ is given by
\beq
\alpha =\frac{\kappa^2 b_0}{6 (1-\Omega^2_0)} (\delta V_* \Omega^2_0
+\delta V  \Omega^4_0).
\label{alpha}
\eeq
The conclusion is  that if $\alpha$ can be neglected, 4d cosmology reduces to the conventional Hubble Law.
The IR distortion of the geometry  proportional to $\delta V$ is precisely what is required to change the curvature by the amount in the IR necessary to remove the mismatch in brane tension. Large perturbations in the IR distory the geometry exactly as is expected from the radion potential.







\begin{thebibliography}{99}



  
  \bibitem{sn1}
{\bf Supernova Search Team} Collaboration, A.~G. Riess {\em et.~al.},
  ``{Observational evidence from supernovae for an accelerating universe and a
  cosmological constant},'' {\em Astron. J.} {\bf 116} (1998) 1009--1038,
  \href{http://xxx.lanl.gov/abs/astro-ph/9805201}{{\tt astro-ph/9805201}}.

\bibitem{sn2}
{\bf Supernova Cosmology Project} Collaboration, S.~Perlmutter {\em et.~al.},
  ``{Measurements of Omega and Lambda from 42 high redshift supernovae},'' {\em
  Astrophys. J.} {\bf 517} (1999) 565--586,
  \href{http://xxx.lanl.gov/abs/astro-ph/9812133}{{\tt astro-ph/9812133}}.
  
  \bibitem{kklt}
S.~Kachru, R.~Kallosh, A.~D. Linde, and S.~P. Trivedi, ``{De Sitter vacua in
  string theory},'' {\em Phys. Rev.} {\bf D68} (2003) 046005,
  \href{http://xxx.lanl.gov/abs/hep-th/0301240}{{\tt hep-th/0301240}}.
  
  \bibitem{fluxcompactification} 
E.~Silverstein, ``{TASI / PiTP / ISS lectures on moduli and microphysics},'' in
  {\em {Progress in string theory. Proceedings, Summer School, TASI 2003,
  Boulder, USA, June 2-27, 2003}}, pp.~381--415, 2004.
\newblock \href{http://xxx.lanl.gov/abs/hep-th/0405068}{{\tt hep-th/0405068}}, 
M.~Grana, ``{Flux compactifications in string theory: A Comprehensive
  review},'' {\em Phys.Rept.} {\bf 423} (2006) 91--158,
  \href{http://xxx.lanl.gov/abs/hep-th/0509003}{{\tt hep-th/0509003}},
M.~R. Douglas and S.~Kachru, ``{Flux compactification},'' {\em Rev.Mod.Phys.}
  {\bf 79} (2007) 733--796, \href{http://xxx.lanl.gov/abs/hep-th/0610102}{{\tt
  hep-th/0610102}},
F.~Denef, ``{Les Houches Lectures on Constructing String Vacua},'' {\em Les
  Houches} {\bf 87} (2008) 483--610,
  \href{http://xxx.lanl.gov/abs/0803.1194}{{\tt 0803.1194}},
F.~Denef, M.~R. Douglas, and S.~Kachru, ``{Physics of String Flux
  Compactifications},'' {\em Ann.Rev.Nucl.Part.Sci.} {\bf 57} (2007) 119--144,
  \href{http://xxx.lanl.gov/abs/hep-th/0701050}{{\tt hep-th/0701050}},
H.~Samtleben, ``{Lectures on Gauged Supergravity and Flux Compactifications},''
  {\em Class. Quant. Grav.} {\bf 25} (2008) 214002,
  \href{http://xxx.lanl.gov/abs/0808.4076}{{\tt 0808.4076}}.
  
  \bibitem{dr}
  U.~H.~Danielsson and T.~Van Riet,
  Int.\ J.\ Mod.\ Phys.\ D {\bf 27}, no. 12, 1830007 (2018)
  doi:10.1142/S0218271818300070
  [arXiv:1804.01120 [hep-th]].
  
  
  \bibitem{also} 
V.~Balasubramanian, P.~Berglund, J.~P. Conlon, and F.~Quevedo, ``{Systematics
  of moduli stabilisation in Calabi-Yau flux compactifications},'' {\em JHEP}
  {\bf 0503} (2005) 007, \href{http://xxx.lanl.gov/abs/hep-th/0502058}{{\tt
  hep-th/0502058}},
O.~Lebedev, H.~P. Nilles, and M.~Ratz, ``{De Sitter vacua from matter
  superpotentials},'' {\em Phys. Lett.} {\bf B636} (2006) 126--131,
  \href{http://xxx.lanl.gov/abs/hep-th/0603047}{{\tt hep-th/0603047}},
C.~P. Burgess, R.~Kallosh, and F.~Quevedo, ``{De Sitter string vacua from
  supersymmetric D terms},'' {\em JHEP} {\bf 10} (2003) 056,
  \href{http://xxx.lanl.gov/abs/hep-th/0309187}{{\tt hep-th/0309187}},
S.~L. Parameswaran and A.~Westphal, ``{de Sitter string vacua from perturbative
  Kahler corrections and consistent D-terms},'' {\em JHEP} {\bf 10} (2006) 079,
  \href{http://xxx.lanl.gov/abs/hep-th/0602253}{{\tt hep-th/0602253}},
A.~Achucarro, B.~de~Carlos, J.~A. Casas, and L.~Doplicher, ``{De Sitter vacua
  from uplifting D-terms in effective supergravities from realistic strings},''
  {\em JHEP} {\bf 06} (2006) 014,
  \href{http://xxx.lanl.gov/abs/hep-th/0601190}{{\tt hep-th/0601190}}.
G.~Villadoro and F.~Zwirner, ``{De-Sitter vacua via consistent D-terms},'' {\em
  Phys. Rev. Lett.} {\bf 95} (2005) 231602,
  \href{http://xxx.lanl.gov/abs/hep-th/0508167}{{\tt hep-th/0508167}},
A.~Saltman and E.~Silverstein, ``{The Scaling of the no scale potential and de
  Sitter model building},'' {\em JHEP} {\bf 11} (2004) 066,
  \href{http://xxx.lanl.gov/abs/hep-th/0402135}{{\tt hep-th/0402135}},
D.~Gallego, M.~C.~D. Marsh, B.~Vercnocke, and T.~Wrase, ``{A New Class of de
  Sitter Vacua in Type IIB Large Volume Compactifications},'' {\em JHEP} {\bf
  10} (2017) 193, \href{http://xxx.lanl.gov/abs/1707.01095}{{\tt 1707.01095}},
A.~Westphal, ``{de Sitter string vacua from Kahler uplifting},'' {\em JHEP}
  {\bf 03} (2007) 102, \href{http://xxx.lanl.gov/abs/hep-th/0611332}{{\tt
  hep-th/0611332}},
S.~L. Parameswaran and A.~Westphal, ``{Consistent de Sitter string vacua from
  Kahler stabilization and D-term uplifting},'' {\em Fortsch. Phys.} {\bf 55}
  (2007) 804--810, \href{http://xxx.lanl.gov/abs/hep-th/0701215}{{\tt
  hep-th/0701215}},
M.~Rummel and A.~Westphal, ``{A sufficient condition for de Sitter vacua in
  type IIB string theory},'' {\em JHEP} {\bf 01} (2012) 020,
  \href{http://xxx.lanl.gov/abs/1107.2115}{{\tt 1107.2115}},
J.~Louis, M.~Rummel, R.~Valandro, and A.~Westphal, ``{Building an explicit de
  Sitter},'' {\em JHEP} {\bf 10} (2012) 163,
  \href{http://xxx.lanl.gov/abs/1208.3208}{{\tt 1208.3208}},
M.~Cicoli, A.~Maharana, F.~Quevedo, and C.~P. Burgess, ``{De Sitter String
  Vacua from Dilaton-dependent Non-perturbative Effects},'' {\em JHEP} {\bf 06}
  (2012) 011, \href{http://xxx.lanl.gov/abs/1203.1750}{{\tt 1203.1750}},
M.~Cicoli, D.~Klevers, S.~Krippendorf, C.~Mayrhofer, F.~Quevedo, and
  R.~Valandro, ``{Explicit de Sitter Flux Vacua for Global String Models with
  Chiral Matter},'' {\em JHEP} {\bf 05} (2014) 001,
  \href{http://xxx.lanl.gov/abs/1312.0014}{{\tt 1312.0014}},
  
  
  
L.~Aparicio, M.~Cicoli, S.~Krippendorf, A.~Maharana, F.~Muia, and F.~Quevedo,
  ``{Sequestered de Sitter String Scenarios: Soft-terms},'' {\em JHEP} {\bf 11}
  (2014) 071, \href{http://xxx.lanl.gov/abs/1409.1931}{{\tt 1409.1931}}.

\bibitem{Cicoli:2015ylx}
M.~Cicoli, F.~Quevedo, and R.~Valandro, ``{De Sitter from T-branes},'' {\em
  JHEP} {\bf 03} (2016) 141, \href{http://xxx.lanl.gov/abs/1512.04558}{{\tt
  1512.04558}}.

\bibitem{Antoniadis:2018hqy}
I.~Antoniadis, Y.~Chen, and G.~K. Leontaris, ``{Perturbative moduli
  stabilisation in type IIB/F-theory framework},''
  \href{http://xxx.lanl.gov/abs/1803.08941}{{\tt 1803.08941}}.

\bibitem{Covi:2008ea}
L.~Covi, M.~Gomez-Reino, C.~Gross, J.~Louis, G.~A. Palma, and C.~A. Scrucca,
  ``{de Sitter vacua in no-scale supergravities and Calabi-Yau string
  models},'' {\em JHEP} {\bf 06} (2008) 057,
  \href{http://xxx.lanl.gov/abs/0804.1073}{{\tt 0804.1073}}.

\bibitem{Covi:2008zu}
L.~Covi, M.~Gomez-Reino, C.~Gross, G.~A. Palma, and C.~A. Scrucca,
  ``{Constructing de Sitter vacua in no-scale string models without
  uplifting},'' {\em JHEP} {\bf 03} (2009) 146,
  \href{http://xxx.lanl.gov/abs/0812.3864}{{\tt 0812.3864}}.

\bibitem{Kachru:2002sk}
S.~Kachru, M.~B. Schulz, P.~K. Tripathy, and S.~P. Trivedi, ``{New
  supersymmetric string compactifications},'' {\em JHEP} {\bf 03} (2003) 061,
  \href{http://xxx.lanl.gov/abs/hep-th/0211182}{{\tt hep-th/0211182}}.

\bibitem{Grana:2006kf}
M.~Grana, R.~Minasian, M.~Petrini, and A.~Tomasiello, ``{A Scan for new N=1
  vacua on twisted tori},'' {\em JHEP} {\bf 05} (2007) 031,
  \href{http://xxx.lanl.gov/abs/hep-th/0609124}{{\tt hep-th/0609124}}.
J.~Blaback, U.~H. Danielsson, D.~Junghans, T.~Van~Riet, T.~Wrase, and
  M.~Zagermann, ``{Smeared versus localised sources in flux
  compactifications},'' {\em JHEP} {\bf 12} (2010) 043,
  \href{http://xxx.lanl.gov/abs/1009.1877}{{\tt 1009.1877}}.
D.~Andriot, J.~Blåbäck, and T.~Van~Riet, ``{Minkowski flux vacua of type II
  supergravities},'' {\em Phys. Rev. Lett.} {\bf 118} (2017), no.~1 011603,
  \href{http://xxx.lanl.gov/abs/1609.00729}{{\tt 1609.00729}},
E.~Palti, G.~Tasinato, and J.~Ward, ``{WEAKLY-coupled IIA Flux
  Compactifications},'' {\em JHEP} {\bf 06} (2008) 084,
  \href{http://xxx.lanl.gov/abs/0804.1248}{{\tt 0804.1248}}.
M.~Davidse, F.~Saueressig, U.~Theis, and S.~Vandoren, ``{Membrane instantons
  and de Sitter vacua},'' {\em JHEP} {\bf 09} (2005) 065,
  \href{http://xxx.lanl.gov/abs/hep-th/0506097}{{\tt hep-th/0506097}},
F.~Saueressig, U.~Theis, and S.~Vandoren, ``{On de Sitter vacua in type IIA
  orientifold compactifications},'' {\em Phys. Lett.} {\bf B633} (2006)
  1F25--128, \href{http://xxx.lanl.gov/abs/hep-th/0506181}{{\tt
  hep-th/0506181}},
J.-P. Derendinger, C.~Kounnas, P.~M. Petropoulos, and F.~Zwirner,
  ``{Superpotentials in IIA compactifications with general fluxes},'' {\em
  Nucl. Phys.} {\bf B715} (2005) 211--233,
  \href{http://xxx.lanl.gov/abs/hep-th/0411276}{{\tt hep-th/0411276}},
O.~DeWolfe, A.~Giryavets, S.~Kachru, and W.~Taylor, ``{Type IIA moduli
  stabilization},'' {\em JHEP} {\bf 07} (2005) 066,
  \href{http://xxx.lanl.gov/abs/hep-th/0505160}{{\tt hep-th/0505160}},
B.~S. Acharya, F.~Benini, and R.~Valandro, ``{Fixing moduli in exact type IIA
  flux vacua},'' {\em JHEP} {\bf 02} (2007) 018,
  \href{http://xxx.lanl.gov/abs/hep-th/0607223}{{\tt hep-th/0607223}},
R.~Kallosh and M.~Soroush, ``{Issues in type IIA uplifting},'' {\em JHEP} {\bf
  06} (2007) 041, \href{http://xxx.lanl.gov/abs/hep-th/0612057}{{\tt
  hep-th/0612057}},
S.~L. Parameswaran, S.~Ramos-Sanchez, and I.~Zavala, ``{On Moduli Stabilisation
  and de Sitter Vacua in MSSM Heterotic Orbifolds},'' {\em JHEP} {\bf 01}
  (2011) 071, \href{http://xxx.lanl.gov/abs/1009.3931}{{\tt 1009.3931}},
M.~Cicoli, S.~de~Alwis, and A.~Westphal, ``{Heterotic Moduli Stabilisation},''
  {\em JHEP} {\bf 10} (2013) 199, \href{http://xxx.lanl.gov/abs/1304.1809}{{\tt
  1304.1809}},
S.~Gukov, S.~Kachru, X.~Liu, and L.~McAllister, ``{Heterotic moduli
  stabilization with fractional Chern-Simons invariants},'' {\em Phys. Rev.}
  {\bf D69} (2004) 086008, \href{http://xxx.lanl.gov/abs/hep-th/0310159}{{\tt
  hep-th/0310159}},
F.~Denef and M.~R. Douglas, ``{Distributions of nonsupersymmetric flux
  vacua},'' {\em JHEP} {\bf 03} (2005) 061,
  \href{http://xxx.lanl.gov/abs/hep-th/0411183}{{\tt hep-th/0411183}}.


  
  
 \bibitem{moritz}
  J.~Moritz, A.~Retolaza and A.~Westphal,
  Phys.\ Rev.\ D {\bf 97}, no. 4, 046010 (2018)
  doi:10.1103/PhysRevD.97.046010
  [arXiv:1707.08678 [hep-th]].
J.~Moritz, A.~Retolaza, and A.~Westphal, ``{Towards de Sitter from 10D},''
  \href{http://xxx.lanl.gov/abs/1707.08678}{{\tt 1707.08678}}.
  
  \bibitem{sav} 
S.~Sethi, ``{Supersymmetry Breaking by Fluxes},''
  \href{http://xxx.lanl.gov/abs/1709.03554}{{\tt 1709.03554}},
S.~R. Green, E.~J. Martinec, C.~Quigley, and S.~Sethi, ``{Constraints on String
  Cosmology},'' {\em Class. Quant. Grav.} {\bf 29} (2012) 075006,
  \href{http://xxx.lanl.gov/abs/1110.0545}{{\tt 1110.0545}},
D.~Kutasov, T.~Maxfield, I.~Melnikov, and S.~Sethi, ``{Constraining de Sitter
  Space in String Theory},'' {\em Phys. Rev. Lett.} {\bf 115} (2015), no.~7
  071305, \href{http://xxx.lanl.gov/abs/1504.00056}{{\tt 1504.00056}}.

  
  
\bibitem{polchinski}
J.~Polchinski, ``{Brane/antibrane dynamics and KKLT stability},''
  \href{http://xxx.lanl.gov/abs/1509.05710}{{\tt 1509.05710}}, 
B.~Michel, E.~Mintun, J.~Polchinski, A.~Puhm, and P.~Saad, ``{Remarks on brane
  and antibrane dynamics},'' {\em JHEP} {\bf 09} (2015) 021,
  \href{http://xxx.lanl.gov/abs/1412.5702}{{\tt 1412.5702}}.
  
  

\bibitem{kkmz} S. Kachru, m.Kim, L. McAllister, M. Zimet"de Sitter Vacua from Ten Dimensions   " \href(https://arxiv.org/abs/1908.04788).

\bibitem{antid31}
  J.~Armas, N.~Nguyen, V.~Niarchos, N.~A.~Obers and T.~Van Riet,
  Phys.\ Rev.\ Lett.\  {\bf 122}, no. 18, 181601 (2019)
  doi:10.1103/PhysRevLett.122.181601
  [arXiv:1812.01067 [hep-th]].
  
\bibitem{antid32}
  Y.~Hamada, A.~Hebecker, G.~Shiu and P.~Soler,
  JHEP {\bf 1904}, 008 (2019)
  doi:10.1007/JHEP04(2019)008
  [arXiv:1812.06097 [hep-th]],
  JHEP {\bf 1906} (2019) 019
  doi:10.1007/JHEP06(2019)019
  [arXiv:1902.01410 [hep-th]].


 


\bibitem{ns5}
R.~C. Myers, ``{Dielectric branes},'' {\em JHEP} {\bf 12} (1999) 022,
  \href{http://xxx.lanl.gov/abs/hep-th/9910053}{{\tt hep-th/9910053}}, 
D.~Cohen-Maldonado, J.~Diaz, and F.~F. Gautason, ``{Polarised antibranes from
  Smarr relations},'' {\em JHEP} {\bf 05} (2016) 175,
  \href{http://xxx.lanl.gov/abs/1603.05678}{{\tt 1603.05678}}.

\bibitem{kpv} S.~Kachru, J.~Pearson, and H.~L. Verlinde, ``{Brane / flux annihilation and the
  string dual of a nonsupersymmetric field theory},'' {\em JHEP} {\bf 06}
  (2002) 021, \href{http://xxx.lanl.gov/abs/hep-th/0112197}{{\tt
  hep-th/0112197}}.

\bibitem{gwkklt} Felix Bruemmer and Arthur Hebecker and Enrico Trincherini, The Throat as a Randall-Sundrum Model with Goldberger-Wise Stabilization,
    \href(hep-th/0510113).
    
    
  
  \bibitem{gkp}
S.~B. Giddings, S.~Kachru, and J.~Polchinski, ``{Hierarchies from fluxes in
  string compactifications},'' {\em Phys.Rev.} {\bf D66} (2002) 106006,
  \href{http://xxx.lanl.gov/abs/hep-th/0105097}{{\tt hep-th/0105097}}.
  
  
\bibitem{ks}
I.~R. Klebanov and M.~J. Strassler, ``{Supergravity and a confining gauge
  theory: Duality cascades and chi SB resolution of naked singularities},''
  {\em JHEP} {\bf 08} (2000) 052,
  \href{http://xxx.lanl.gov/abs/hep-th/0007191}{{\tt hep-th/0007191}}.


\bibitem{verlinde}
H.~L.~Verlinde,
  Nucl.\ Phys.\ B {\bf 580}, 264 (2000)
  doi:10.1016/S0550-3213(00)00224-8
  [hep-th/9906182].
  
\bibitem{rs1} 
  L.~Randall and R.~Sundrum,
  Phys.\ Rev.\ Lett.\  {\bf 83}, 3370 (1999)
  doi:10.1103/PhysRevLett.83.3370
  [hep-ph/9905221].
  
\bibitem{rs2}
L.~Randall and R.~Sundrum, ``{An Alternative to compactification},'' {\em Phys.
  Rev. Lett.} {\bf 83} (1999) 4690--4693,
  \href{http://xxx.lanl.gov/abs/hep-th/9906064}{{\tt hep-th/9906064}}.
  
  
  
  \bibitem{kr}
  A.~Karch and L.~Randall,
  Phys.\ Rev.\ Lett.\  {\bf 87}, 061601 (2001)
  doi:10.1103/PhysRevLett.87.061601
  [hep-th/0105108].
  
  \bibitem{kaloper}
N.~Kaloper, ``Bent domain walls as brane-worlds,''
  \href{http://xxx.lanl.gov/abs/hep-th/9905210}{{\tt hep-th/9905210}}.
  
  
  \bibitem{dfk}
O.~DeWolfe, D.~Z. Freedman, S.~S. Gubser, and A.~Karch, ``Modeling the fifth
  dimension with scalars and gravity,''
  \href{http://xxx.lanl.gov/abs/hep-th/9909134}{{\tt hep-th/9909134}}.
  
 
\bibitem{cvetic1}
M.~Cvetic, S.~Griffies, and H.~H. Soleng, ``Local and global gravitational
  aspects of domain wall space-times,'' {\em Phys. Rev.} {\bf D48} (1993)
  2613--2634, \href{http://xxx.lanl.gov/abs/gr-qc/9306005}{{\tt
  gr-qc/9306005}}.

\bibitem{cvetic2}
M.~Cvetic and J.~Wang, ``Vacuum domain walls in D-dimensions: Local and global
  space-time structure,'' {\em Phys. Rev.} {\bf D61} (2000) 124020,
  \href{http://xxx.lanl.gov/abs/hep-th/9912187}{{\tt hep-th/9912187}}.


  \bibitem{kimsquared}
H.~B. Kim and H.~D. Kim, ``Inflation and gauge hierarchy in Randall-Sundrum
  compactification,'' \href{http://xxx.lanl.gov/abs/hep-th/9909053}{{\tt
  hep-th/9909053}}.

\bibitem{nihei}
T.~Nihei, ``Inflation in the five-dimensional universe with an orbifold extra
  dimension,'' \href{http://xxx.lanl.gov/abs/hep-ph/9905487}{{\tt
  hep-ph/9905487}}.
  
 \bibitem{kr3} A. Karch and L. Randall, in progress.
  
  

\bibitem{bena}
I.~Bena, M.~Grana, and N.~Halmagyi, ``{On the Existence of Meta-stable Vacua in
  Klebanov-Strassler},'' {\em JHEP} {\bf 1009} (2010) 087,
  \href{http://xxx.lanl.gov/abs/0912.3519}{{\tt 0912.3519}},

I.~Bena, A.~Buchel, and O.~J. Dias, ``{Horizons cannot save the Landscape},''
  {\em Phys.Rev.} {\bf D87} (2013) 063012,
  \href{http://xxx.lanl.gov/abs/1212.5162}{{\tt 1212.5162}},
I.~Bena, M.~Grana, S.~Kuperstein, and S.~Massai, ``{Tachyonic Anti-M2
  Branes},'' {\em JHEP} {\bf 1406} (2014) 173,
  \href{http://xxx.lanl.gov/abs/1402.2294}{{\tt 1402.2294}},
I.~Bena and S.~Kuperstein, ``{Brane polarization is no cure for tachyons},''
  {\em JHEP} {\bf 09} (2015) 112,
  \href{http://xxx.lanl.gov/abs/1504.00656}{{\tt 1504.00656}},
I.~Bena, J.~Blaback, and D.~Turton, ``{Loop corrections to the antibrane
  potential},'' {\em JHEP} {\bf 07} (2016) 132,
  \href{http://xxx.lanl.gov/abs/1602.05959}{{\tt 1602.05959}}
  
  \bibitem{riet}
U.~H. Danielsson, S.~S. Haque, G.~Shiu, and T.~Van~Riet, ``{Towards Classical
  de Sitter Solutions in String Theory},'' {\em JHEP} {\bf 09} (2009) 114,
  \href{http://xxx.lanl.gov/abs/0907.2041}{{\tt 0907.2041}}, 
J.~Blaback, U.~H. Danielsson, and T.~Van~Riet, ``{Resolving anti-brane
  singularities through time-dependence},'' {\em JHEP} {\bf 02} (2013) 061,
  \href{http://xxx.lanl.gov/abs/1202.1132}{{\tt 1202.1132}},
U.~H. Danielsson and T.~Van~Riet, {Fatal attraction: more on decaying
  anti-branes}, {\em JHEP} {\bf 03} (2015) 087,
  \href{http://xxx.lanl.gov/abs/1410.8476}{{\tt 1410.8476}},
U.~H. Danielsson, F.~F. Gautason, and T.~Van~Riet, ``{Unstoppable brane-flux
  decay of $ \overline{\mathrm{D}6} $ branes},'' {\em JHEP} {\bf 03} (2017)
  141, \href{http://xxx.lanl.gov/abs/1609.06529}{{\tt 1609.06529}},
D.~Cohen-Maldonado, J.~Diaz, T.~Van~Riet, and B.~Vercnocke, ``{Observations on
  fluxes near anti-branes},'' {\em JHEP} {\bf 01} (2016) 126,
  \href{http://xxx.lanl.gov/abs/1507.01022}{{\tt 1507.01022}}, 
U.~H. Danielsson, ``{Perturbative decay of anti-branes in flux backgrounds due
  to space time instabilities},''
  \href{http://xxx.lanl.gov/abs/1502.01234}{{\tt 1502.01234}}.
  
\bibitem{gdw} 
  O.~DeWolfe and S.~B.~Giddings,
  Phys.\ Rev.\ D {\bf 67}, 066008 (2003)
  doi:10.1103/PhysRevD.67.066008
  [hep-th/0208123].
  
  \bibitem{Douglas}
  M. Douglas, Effective potential and warp factor dynamics,
   ISSN={1029-8479},
   url={http://dx.doi.org/10.1007/JHEP03(2010)071},
   DOI={10.1007/jhep03(2010)071}.



\bibitem{gw} 
  W.~D.~Goldberger and M.~B.~Wise,
  Phys.\ Rev.\ Lett.\  {\bf 83}, 4922 (1999)
  doi:10.1103/PhysRevLett.83.4922
  [hep-ph/9907447].
  
  \bibitem{43}
  C.~Csaki, H.~Ooguri, Y.~Oz and J.~Terning,
  JHEP {\bf 9901}, 017 (1999)
  doi:10.1088/1126-6708/1999/01/017
  [hep-th/9806021].
  
  \bibitem{44}
  R.~de Mello Koch, A.~Jevicki, M.~Mihailescu and J.~P.~Nunes,
  Phys.\ Rev.\ D {\bf 58} (1998) 105009
  doi:10.1103/PhysRevD.58.105009
  [hep-th/9806125].
  
\bibitem{rz} 
  R.~Rattazzi and A.~Zaffaroni,
  JHEP {\bf 0104}, 021 (2001)
  doi:10.1088/1126-6708/2001/04/021
  [hep-th/0012248].
  
\bibitem{apr} 
  N.~Arkani-Hamed, M.~Porrati and L.~Randall,
  JHEP {\bf 0108}, 017 (2001)
  doi:10.1088/1126-6708/2001/08/017
  [hep-th/0012148].
  
  
  \bibitem{cgrt}
  C.~Csaki, M.~Graesser, L.~Randall and J.~Terning,
  Phys.\ Rev.\ D {\bf 62}, 045015 (2000)
  doi:10.1103/PhysRevD.62.045015
  [hep-ph/9911406].
  
  \bibitem{numericalks} Berg, Marcus and Haack, Michael and Mück, Wolfgang,
  {Glueballs vs. gluinoballs: Fluctuation spectra in non-AdS/non-CFT},
   volume={789},
   ISSN={0550-3213},
   url={http://dx.doi.org/10.1016/j.nuclphysb.2007.07.012},
   DOI={10.1016/j.nuclphysb.2007.07.012},
   number={1-2},
   journal={Nuclear Physics B},
   publisher={Elsevier BV}.

  
 \bibitem{numericalks2}
  I.~Gordeli and D.~Melnikov,
  JHEP {\bf 1108}, 082 (2011)
  doi:10.1007/JHEP08(2011)082
  [arXiv:0912.5517 [hep-th]].
 

\bibitem{numericalks1} Gubser, Steven S and Herzog, Christopher P and Klebanov, Igor R,
Symmetry Breaking and Axionic Strings in the Warped Deformed Conifold,
   volume={2004},
   ISSN={1029-8479},
   url={http://dx.doi.org/10.1088/1126-6708/2004/09/036},
   DOI={10.1088/1126-6708/2004/09/036},
   number={09},
   journal={Journal of High Energy Physics},
   publisher={Springer Nature},
 Variations on the warped deformed conifold,
   volume={5},
   ISSN={1631-0705},
   url={http://dx.doi.org/10.1016/j.crhy.2004.10.003},
   DOI={10.1016/j.crhy.2004.10.003},
   number={9-10},
   journal={Comptes Rendus Physique},
   publisher={Elsevier BV},
   author={Gubser, Steven S. and Herzog, Christopher P. and Klebanov, Igor R.}.


  \bibitem{dst}
  M.~R.~Douglas, J.~Shelton and G.~Torroba,
  arXiv:0704.4001 [hep-th].
  
  \bibitem{dt} 
  M.~R.~Douglas and G.~Torroba,
  JHEP {\bf 0905}, 013 (2009)
  doi:10.1088/1126-6708/2009/05/013
  [arXiv:0805.3700 [hep-th]].
  
  \bibitem{cosmology}   
  P.~Creminelli, A.~Nicolis and R.~Rattazzi,
  ``Holography and the electroweak phase transition,''
  JHEP {\bf 0203}, 051 (2002)
  [arXiv:hep-th/0107141],
  L.~Randall and G.~Servant,
  ``Gravitational Waves from Warped Spacetime,''
  JHEP {\bf 0705} (2007) 054
  [arXiv:hep-ph/0607158],
  G.~Nardini, M.~Quiros and A.~Wulzer,
  arXiv:0706.3388 [hep-ph].
  
  \bibitem{recentcosmo}
  K.~Agashe, P.~Du, M.~Ekhterachian, S.~Kumar and R.~Sundrum,
  arXiv:1910.06238 [hep-ph].
  T.~Konstandin and G.~Servant,
  JCAP {\bf 1112}, 009 (2011)
  doi:10.1088/1475-7516/2011/12/009
  [arXiv:1104.4791 [hep-ph]].
  P.~Baratella, A.~Pomarol and F.~Rompineve,
  JHEP {\bf 1903} (2019) 100
  doi:10.1007/JHEP03(2019)100
  [arXiv:1812.06996 [hep-ph]].


  
  
  \bibitem{jmr}
  B.~Hassanain, J.~March-Russell and M.~Schvellinger,
  JHEP {\bf 0710}, 089 (2007)
  doi:10.1088/1126-6708/2007/10/089
  [arXiv:0708.2060 [hep-th]].
  
  \bibitem{benedict}
  B.~von Harling, A.~Hebecker and T.~Noguchi,
  JHEP {\bf 0711}, 042 (2007)
  doi:10.1088/1126-6708/2007/11/042
  [arXiv:0705.3648 [hep-th]].
  
 
  
\bibitem{chacko} 
  Z.~Chacko, R.~K.~Mishra and D.~Stolarski,
  JHEP {\bf 1309}, 121 (2013)
  doi:10.1007/JHEP09(2013)121
  [arXiv:1304.1795 [hep-ph]].
  
 
  
\bibitem{raman} 
  P.~Agrawal and R.~Sundrum,
  JHEP {\bf 1705}, 144 (2017)
  doi:10.1007/JHEP05(2017)144
  [arXiv:1611.07021 [hep-th]].
  
  \bibitem{csaba}
  B.~Bellazzini, C.~Csaki, J.~Hubisz, J.~Serra and J.~Terning,
  Eur.\ Phys.\ J.\ C {\bf 74}, 2790 (2014)
  doi:10.1140/epjc/s10052-014-2790-x
  [arXiv:1305.3919 [hep-th]].
 
  
\bibitem{benarunning1} 
  I.~Bena, E.~Dudas, M.~Graña and S.~Lüst,
  Fortsch.\ Phys.\  {\bf 67}, no. 1-2, 1800100 (2019)
  [Fortsch.\ Phys.\  {\bf 2018}, 1800100]
  doi:10.1002/prop.201800100
  [arXiv:1809.06861 [hep-th]],
  R.~Blumenhagen, D.~Kläwer and L.~Schlechter,
  JHEP {\bf 1905}, 152 (2019)
  doi:10.1007/JHEP05(2019)152
  [arXiv:1902.07724 [hep-th]].
  
\bibitem{buchel} 
  A.~Buchel,
  JHEP {\bf 1901}, 207 (2019)
  doi:10.1007/JHEP01(2019)207
  [arXiv:1809.08484 [hep-th]].
  
   
\bibitem{benarunning2} 
  I.~Bena, A.~Buchel and S.~Lüst,
  arXiv:1910.08094 [hep-th].
  
  \bibitem{mcg}
  \cite{McGuirk:2009xx}
  P.~McGuirk, G.~Shiu and Y.~Sumitomo,
  Nucl.\ Phys.\ B {\bf 842}, 383 (2011)
  doi:10.1016/j.nuclphysb.2010.09.008
  [arXiv:0910.4581 [hep-th]].
  
  
\bibitem{rs0} 
  L.~Randall and R.~Sundrum,
  Nucl.\ Phys.\ B {\bf 557}, 79 (1999)
  doi:10.1016/S0550-3213(99)00359-4
  [hep-th/9810155].
  
\bibitem{lmt} 
  G.~F.~Giudice, M.~A.~Luty, H.~Murayama and R.~Rattazzi,
  JHEP {\bf 9812}, 027 (1998)
  doi:10.1088/1126-6708/1998/12/027
  [hep-ph/9810442].
  
  \bibitem{nilles}
K.~Choi, A.~Falkowski, H.~P. Nilles, and M.~Olechowski, ``{Soft supersymmetry
  breaking in KKLT flux compactification},'' {\em Nucl. Phys.} {\bf B718}
  (2005) 113--133, \href{http://xxx.lanl.gov/abs/hep-th/0503216}{{\tt
  hep-th/0503216}}.
  
  \bibitem{cfno}Choi, Kiwoon and Jeong, Kwang-Sik and Okumura, Ken-ichi,  Phenomenology of mixed modulus-anomaly mediation in fluxed string compactifications and brane models,
   volume={2005},
   ISSN={1029-8479},
   url={http://dx.doi.org/10.1088/1126-6708/2005/09/039},
   DOI={10.1088/1126-6708/2005/09/039},
   number={09},
   journal={Journal of High Energy Physics},
   publisher={Springer Nature}.
   
   
   \bibitem{goldstino}
M.~Bertolini, D.~Musso, I.~Papadimitriou, and H.~Raj, ``{A goldstino at the
  bottom of the cascade},'' {\em JHEP} {\bf 11} (2015) 184,
  \href{http://xxx.lanl.gov/abs/1509.03594}{{\tt 1509.03594}},
C.~Krishnan, H.~Raj, and P.~N. Bala~Subramanian, ``{On the KKLT Goldstino},''
  \href{http://xxx.lanl.gov/abs/1803.04905}{{\tt 1803.04905}}.
  
  
   \bibitem{sundrumkklt}
  R.~Sundrum,
  JHEP {\bf 1101}, 062 (2011)
  doi:10.1007/JHEP01(2011)062
  [arXiv:0909.5430 [hep-th]].
   
   
   \bibitem{gp} Gherghetta, Tony and Pomarol, Alex, Gherghetta,
   The standard model partly supersymmetric,
   volume={67},
   ISSN={1089-4918},
   url={http://dx.doi.org/10.1103/PhysRevD.67.085018},
   DOI={10.1103/physrevd.67.085018},
   number={8},
   journal={Physical Review D},
   publisher={American Physical Society (APS).
   
   
     
  \bibitem{ramankklt}  
  R.~Sundrum,
  JHEP {\bf 1101}, 062 (2011)
  doi:10.1007/JHEP01(2011)062
  [arXiv:0909.5430 [hep-th]].
  
  \bibitem{split1}
N.~Arkani-Hamed and S.~Dimopoulos,
  ``Supersymmetric unification without low energy supersymmetry 
and  signatures
  for fine-tuning at the LHC,''
  JHEP {\bf 0506}, 073 (2005)
  [arXiv:hep-th/0405159].

\bibitem{split2} 
G.~F.~Giudice and A.~Romanino,
  ``Split supersymmetry,''
  Nucl.\ Phys.\  B {\bf 699}, 65 (2004)
  [Erratum-ibid.\  B {\bf 706}, 65 (2005)]
  [arXiv:hep-ph/0406088].
 
\bibitem{split3}
 N.~Arkani-Hamed, S.~Dimopoulos, G.~F.~Giudice and A.~Romanino,
  ``Aspects of split supersymmetry,''
  Nucl.\ Phys.\  B {\bf 709}, 3 (2005)
  [arXiv:hep-ph/0409232].




\bibitem{moreminimal}
  A.~G.~Cohen, D.~B.~Kaplan and A.~E.~Nelson,
  ``The more minimal supersymmetric standard model,''
  Phys.\ Lett.\  B {\bf 388}, 588 (1996)
  [arXiv:hep-ph/9607394].



   
   \bibitem{crt} 
  C.~Csaki, L.~Randall and J.~Terning,
  Phys.\ Rev.\ D {\bf 86}, 075009 (2012)
  doi:10.1103/PhysRevD.86.075009
  [arXiv:1201.1293 [hep-ph]].

  \bibitem{lutysundrum}
  M.~A.~Luty and R.~Sundrum,
  Phys.\ Rev.\ D {\bf 64}, 065012 (2001)
  doi:10.1103/PhysRevD.64.065012
  [hep-th/0012158].

\bibitem{lutysundrum2}
  M.~A.~Luty and R.~Sundrum,
  Phys.\ Rev.\ D {\bf 62} (2000) 035008
  doi:10.1103/PhysRevD.62.035008
  [hep-th/9910202].

  
  \bibitem{strassler4} Matthew J. Strassler, Non-Supersymmetric Theories with Light Scalar Fields and Large Hierarchies,
    year={2003},
    eprint={hep-th/0309122},
    archivePrefix={arXiv},
    primaryClass={hep-th}.

  
  
  
  \bibitem{kt4} Kachru, Shamit and Simić, Dušan and Trivedi, Sandip P.,
  Stable non-supersymmetric throats in string theory,
   volume={2010},
   ISSN={1029-8479},
   url={http://dx.doi.org/10.1007/JHEP05(2010)067},
   DOI={10.1007/jhep05(2010)067},
   number={5},
   journal={Journal of High Energy Physics},
   publisher={Springer Nature}.
   
   \bibitem{landscape1}
L.~Susskind, ``{The Anthropic landscape of string theory},''
  \href{http://xxx.lanl.gov/abs/hep-th/0302219}{{\tt hep-th/0302219}}.

\bibitem{landscape2}
A.~N. Schellekens, ``{The String Theory Landscape},'' {\em Adv. Ser. Direct.
  High Energy Phys.} {\bf 22} (2015) 155--217.



\bibitem{landscape3}
T.~D. Brennan, F.~Carta, and C.~Vafa, ``{The String Landscape, the Swampland,
  and the Missing Corner},'' \href{http://xxx.lanl.gov/abs/1711.00864}{{\tt
  1711.00864}}.

   
   
   
   \bibitem{largefieldswampland}
  H.~Ooguri and C.~Vafa,
  Nucl.\ Phys.\ B {\bf 766}, 21 (2007)
  doi:10.1016/j.nuclphysb.2006.10.033
  [hep-th/0605264].
  
  \bibitem{swampland} 
C.~Vafa, ``{The String landscape and the swampland},''
  \href{http://xxx.lanl.gov/abs/hep-th/0509212}{{\tt hep-th/0509212}},
N.~Arkani-Hamed, L.~Motl, A.~Nicolis, and C.~Vafa, ``{The String landscape,
  black holes and gravity as the weakest force},'' {\em JHEP} {\bf 06} (2007)
  060, \href{http://xxx.lanl.gov/abs/hep-th/0601001}{{\tt hep-th/0601001}},
M.~Petrini, G.~Solard, and T.~Van~Riet, ``{AdS vacua with scale separation from
  IIB supergravity},'' {\em JHEP} {\bf 11} (2013) 010,
  \href{http://xxx.lanl.gov/abs/1308.1265}{{\tt 1308.1265}},
  \bibitem{Danielsson:2016rmq}
U.~H. Danielsson, G.~Dibitetto, and S.~C. Vargas, ``{Universal isolation in the
  AdS landscape},'' {\em Phys. Rev.} {\bf D94} (2016), no.~12 126002,
  \href{http://xxx.lanl.gov/abs/1605.09289}{{\tt 1605.09289}},
U.~H. Danielsson, G.~Dibitetto, and S.~C. Vargas, ``{A swamp of non-SUSY
  vacua},'' {\em JHEP} {\bf 11} (2017) 152,
  \href{http://xxx.lanl.gov/abs/1708.03293}{{\tt 1708.03293}},
E.~Kiritsis, F.~Nitti, and L.~Silva~Pimenta, ``{Exotic RG Flows from
  Holography},'' {\em Fortsch. Phys.} {\bf 65} (2017), no.~2 1600120,
  \href{http://xxx.lanl.gov/abs/1611.05493}{{\tt 1611.05493}},
H.~Ooguri and C.~Vafa, ``{Non-supersymmetric AdS and the Swampland},''
  \href{http://xxx.lanl.gov/abs/1610.01533}{{\tt 1610.01533}},
U.~Danielsson and G.~Dibitetto, ``{Fate of stringy AdS vacua and the weak
  gravity conjecture},'' {\em Phys. Rev.} {\bf D96} (2017), no.~2 026020,
  \href{http://xxx.lanl.gov/abs/1611.01395}{{\tt 1611.01395}}.


  
   
   

  
\bibitem{mn}
J.~M. Maldacena and C.~Nunez, ``{Supergravity description of field theories on
  curved manifolds and a no go theorem},'' {\em Int. J. Mod. Phys.} {\bf A16}
  (2001) 822--855, \href{http://xxx.lanl.gov/abs/hep-th/0007018}{{\tt
  hep-th/0007018}}. [,182(2000)].


\bibitem{underwood}
  S.~Das, S.~S.~Haque and B.~Underwood,
  Phys.\ Rev.\ D {\bf 100}, no. 4, 046013 (2019)
  doi:10.1103/PhysRevD.100.046013
  [arXiv:1905.05864 [hep-th]].
  
  
\bibitem{adk}
  N.~Arkani-Hamed, S.~Dimopoulos, N.~Kaloper and J.~March-Russell,
  Nucl.\ Phys.\ B {\bf 567} (2000) 189
  doi:10.1016/S0550-3213(99)00667-7
  [hep-ph/9903224].
  
  \bibitem{sakai}
  M.~Eto, N.~Maru and N.~Sakai,
  Phys.\ Rev.\ D {\bf 70}, 086002 (2004)
  doi:10.1103/PhysRevD.70.086002
  [hep-th/0403009].
  
  
\bibitem{dewk}
  O.~DeWolfe, S.~Kachru and M.~Mulligan,
  Phys.\ Rev.\ D {\bf 77}, 065011 (2008)
  doi:10.1103/PhysRevD.77.065011
  [arXiv:0801.1520 [hep-th]].
  
  \bibitem{bht}
  F.~Brummer, A.~Hebecker and M.~Trapletti,
  Nucl.\ Phys.\ B {\bf 755}, 186 (2006)
  doi:10.1016/j.nuclphysb.2006.08.002
  [hep-th/0605232].



  
   
  
 

 \bibitem{Dterm}
 Lisa Randall, Talk SUSY '02, $http://www-library.desy.de/preparch/desy/proc/proc02-02/html-pages/susy02_program.html$,
  P.Batra, A.Delgado, D.E.Kaplan and T.M.P.Tait,
  JHEP {\bf 0402}, 043 (2004)
  doi:10.1088/1126-6708/2004/02/043
  [hep-ph/0309149], 
  JHEP {\bf 0606}, 034 (2006)
  doi:10.1088/1126-6708/2006/06/034
  [hep-ph/0409127].


\bibitem{kklmmt}
  S.~Kachru, R.~Kallosh, A.~D.~Linde, J.~M.~Maldacena, L.~P.~McAllister and S.~P.~Trivedi,
  JCAP {\bf 0310} (2003) 013
  doi:10.1088/1475-7516/2003/10/013
  [hep-th/0308055].


  
  
  




\bibitem{gibbons}
G.~W. Gibbons and C.~M. Hull, ``{De Sitter space from warped supergravity
  solutions,}'' \href{http://xxx.lanl.gov/abs/hep-th/0111072}
  {{\tt hep-th/0111072}}.
  
  \bibitem{ugawa}
M.~Minamitsuji and K.~Uzawa, ``{Warped de Sitter compactifications},'' {\em
  JHEP} {\bf 01} (2012) 142, \href{http://xxx.lanl.gov/abs/1103.5326}{{\tt
  1103.5326}}.
  
  
  
  \bibitem{mathpapers} Marc Troyanov,
 Trans. Amer. Math. Soc. 324 (1991), 793-821,
DOI: https://doi.org/10.1090/S0002-9947-1991-1005085-9
MathSciNet review: 1005085.










\bibitem{evaneg}
E.~Silverstein, ``{(A)dS backgrounds from asymmetric orientifolds},'' in {\em
  {Strings 2001: International Conference Mumbai, India, January 5-10, 2001}},
  2001.
\newblock \href{http://xxx.lanl.gov/abs/hep-th/0106209}{{\tt hep-th/0106209}},
A.~Maloney, E.~Silverstein, and A.~Strominger, ``{De Sitter space in
  noncritical string theory},'' in {\em {Workshop on Conference on the Future
  of Theoretical Physics and Cosmology in Honor of Steven Hawking's 60th
  Birthday Cambridge, England, January 7-10, 2002}}, pp.~570--591, 2002.
\newblock \href{http://xxx.lanl.gov/abs/hep-th/0205316}{{\tt hep-th/0205316}}, 
E.~Silverstein, ``{Simple de Sitter Solutions},'' {\em Phys. Rev.} {\bf D77}
  (2008) 106006, \href{http://xxx.lanl.gov/abs/0712.1196}{{\tt 0712.1196}}. 
 \bibitem{kt} 
  S.~Kachru and S.~P.~Trivedi,
  Fortsch.\ Phys.\  {\bf 67}, no. 1-2, 1800086 (2019)
  doi:10.1002/prop.201800086
  [arXiv:1808.08971 [hep-th]].
  
  \bibitem{douglasneg}
  M.~R.~Douglas and R.~Kallosh,
  `Compactification on negatively curved manifolds,''
  JHEP {\bf 1006}, 004 (2010)
  doi:10.1007/JHEP06(2010)004
  [arXiv:1001.4008 [hep-th]].
  
 





  
 












  
%


 

\bibitem{McGuirk:2009xx}
P.~McGuirk, G.~Shiu, and Y.~Sumitomo, ``{Non-supersymmetric infrared
  perturbations to the warped deformed conifold},'' {\em Nucl.Phys.} {\bf B842}
  (2011) 383--413, \href{http://xxx.lanl.gov/abs/0910.4581}{{\tt 0910.4581}}.




}
\end{thebibliography}
\end{document}